\documentclass[useAMS,usenatbib]{mn2e}
\usepackage{graphicx}
\usepackage{longtable}
\usepackage{amssymb,amsmath}

\newcommand{\msol}{M$_{\odot}$}
\newcommand{\msolyr}{M$_{\odot}$\,yr$^{-1}$}

\newcommand{\ha}{H$\alpha$}
\newcommand{\hb}{H$\beta$}

\newcommand{\heiiwr}{He\,{\small II}~$\lambda~4686$}

\newcommand{\niiiwr}{N\,{\small III}~$\lambda\lambda~4634/42$}

\newcommand{\civheib}{He\,{\small I}~$\lambda~4713$}
\newcommand{\niineb}{[N\,{\small II}]~$\lambda~5755$}
\newcommand{\civwrr}{C\,{\small IV}~$\lambda\lambda~5802/12$}
\newcommand{\civheir}{He\,{\small I}~$\lambda~5876$}
\newcommand{\ciiiwrb}{C\,{\small III}~$\lambda~4650$}
\newcommand{\civwrb}{C\,{\small IV}~$\lambda~4658$}

\newcommand{\oviwr}{O\,{\small VI}~$\lambda\lambda~3811/34$}

\newcommand{\ciii}{C\,{\small III}}
\newcommand{\civ}{C\,{\small IV}}

\newcommand{\niii}{N\,{\small III}}
\newcommand{\niv}{N\,{\small IV}}
\newcommand{\nv}{N\,{\small V}}

\newcommand{\fwc}{$f_{\rm WC}$}
\newcommand{\fwn}{$f_{\rm WN}$}

\title[Wolf-Rayet stars in M81]{Wolf-Rayet stars in M81: Detection and 
Characterization using GTC/OSIRIS spectra and HST/ACS images}
\author[V. M. A. G\'omez-Gonz\'alez, Y. D. Mayya and D. Rosa-Gonz\'alez]
{V. M. A. G\'omez-Gonz\'alez, Y. D. Mayya and D. Rosa-Gonz\'alez\\
Instituto Nacional de Astrof{\'\i}sica, \'Optica y Electr\'onica, Luis Enrique
Erro 1, Tonantzintla 72840, Puebla, Mexico\\
Email: maico, ydm, danrosa~@inaoep.mx}

\begin{document}

\date{Accepted for publication in MNRAS (\today)}

\maketitle

\begin{abstract}

We here report the properties of Wolf-Rayet (W-R) stars in 14 locations
in the nearby spiral galaxy M81.
These locations were found serendipitously while analysing the slit spectra
of a sample of $\sim150$ star-forming complexes, taken using the
long-slit and Multi-Object spectroscopic modes of the OSIRIS instrument
at the 10.4-m Gran Telescopio Canarias. Colours and magnitudes of the identified
point sources in the Hubble Space Telescope images compare
well with those of individual W-R stars in the Milky Way.
Using templates of
individual W-R stars, we infer that the objects responsible for the observed
W-R features are single stars in 12 locations, comprising of 3 WNLs, 3
WNEs, 2 WCEs and 4 transitional WN/C types. In diagrams involving bump
luminosities and the width of the
bumps, the W-R stars of the same sub-class group together, with the
transitional stars occupying locations intermediate between the
WNE and WCE groups, as expected from the evolutionary models.
However, the observed number of 4 transitional stars out of our sample of 14 is statistically high as compared to the
4\% expected in stellar evolutionary models.

\end{abstract}


\section{Introduction}

The mere detection of Wolf-Rayet (W-R) stars reveals recent burst
of high-mass star formation in galaxies \citep[]{1998Schaerer}. 
This is because the W-R stars are the short-lived descendants of massive 
O-type stars \citep[$>$25 $\mathrm{M}{}_{\odot}$ at solar metallicity 
($\mathrm{Z}{}_{\odot}$);][]{1976Conti, 2003Meynet,2005Meynet}, spending $\sim$10\% of their $\lesssim$5~Myr lifetime as a W-R star. During the W-R phase, the most massive of them pass through nitrogen-rich (WN), and then carbon-rich (WC) 
sequence and finally explode as a supernova.
W-R stars are characterized by heavy 
mass loss \citep[$\sim10^{-5}-10^{-4}$~\msolyr;] [and references therein]{2007Crowther} 
through powerful stellar winds.
The lost mass not only deposits energy, but also freshly processed elements
to the interstellar medium (ISM), which makes these stars important in the
study of the dynamics as well as the chemical evolution of galaxies \citep{2000Hillier}.
W-R stars are also considered as progenitors of long-period $\gamma$-ray bursts
\citep{2006Woosley}. Alternatively, W-R stars can also be formed in relatively lower-mass stars when they
are in a binary system \citep[e. g.][]{2011Vanbeveren}.

In the optical range of the spectrum, the presence of W-R stars can
be easily inferred by their unique, strong, broad emission features,
better known in the literature as W-R bumps. These bumps are mostly composed 
of blended broad emission lines of He, N and C in various ionization states. 
The so called ``blue bump'' around 4600--4700~\AA\  
is the characteristic feature of all W-R stars. Nitrogen sequence (WN subtype) 
stars contain essentially only the ``blue bump'' composed of He (\heiiwr) and N (\niiiwr) 
lines, both being products of H-burning via the CNO cycle. On the other hand, 
the ``red bump'' around 5750--5850~\AA\ occurs essentially only in stars of the carbon (WC) 
and oxygen sequence (WO). These stars have strong emission lines of He (\heiiwr) and C (\ciiiwrb/\civwrb\ 
and \civwrr), with WO stars in addition having lines of O (\oviwr), all these being the products of He-burning via the
triple-$\alpha$ process \citep[]{2007Crowther}.
Hydrogen lines are normally absent in W-R stars, but there is also a 
subset of W-R stars that show hydrogen lines in emission or absorption. 
These stars are H-burning and share some similarities with Of-type stars \citep[]{1995Crowther,2007Crowther}.

Most of our knowledge on W-R stars is based on the study of individual stars 
in the Milky Way (MW) \citep[][see also the web page by Paul
Crowther\footnote{http://www.pacrowther.staff.shef.ac.uk/WRcat/} for an up to 
date catalogue]{2001Hucht}, the Small Magellanic Cloud \citep[SMC:][]{2003Massey} and 
Large Magellanic Cloud \citep[LMC:][]{1999Breysacher}. 
Recent advances in observational techniques, especially the 
wide-field narrow-band imaging capabilities on large telescopes accompanied by
multi-object spectroscopy (MOS), have allowed studies of W-R stars in nearby
galaxies such as M33 \citep{2011Neugent}, M31 \citep{2012Neugent}, M83 
\citep{2005Hadfield}, NGC~7793 \citep{2010Bibby} and M101 \citep{2013Shara}. 
Also large numbers of W-R stars have been detected in 
the integrated spectra of regions in star-forming galaxies 
\citep[e. g.][]{1996Terlevich, 2005Hadfield, 2013Kehrig, 2014Miralles}.
The giant spiral M81 has not yet been a target for a systematic search for
W-R stars in spite of it being at only a distance of 3.63~Mpc
\citep[$m - M = 27.8$;][]{1994Freedman}. 
Until now there is report of only 6 W-R detections in the entire galaxy \citep{2012Patterson}.
We here report the properties of 14 W-R detections, 13 of them new detections, 
all discovered serendipitously
in slit spectra passing through star-forming regions in the disk of M81 using the 
long-slit and MOS configurations of the OSIRIS\footnote{OSIRIS (Optical System 
for Imaging and low-Intermediate-Resolution Integrated Spectroscopy; 
http://www.gtc.iac.es/instruments/osiris/osiris.php) is an imager and spectrograph 
for the optical wavelength range, located in the Nasmyth-B focus of GTC.} 
instrument at the 10.4-m Gran Telescopio Canarias (GTC)\footnote{Gran Telescopio 
Canarias is a Spanish initiative with the participation of
Mexico and the US University of Florida, and is installed at the Roque de
los Muchachos in the island of La Palma. This work is based on the proposals
GTC-10A, GTC-12B and GTC-14A, all using Mexican share of the GTC time.}.
The MOS data presented here are among the first observations using this mode
at the GTC.

This paper is structured as follows: the strategy used for the search of W-R stars,
as well as the data used in this work and their reduction, are described in \S2.
In \S3 we present the technique used to classify the W-R stars.  
Relations between the obtained parameters for different W-R subtypes in
M81 are discussed in the context of currently accepted models for the formation
of different W-R subtypes in \S4. Results are summarized in \S5.

\section{W-R Sample and observational data}

\subsection{The Sample and Spectroscopic Observations}

The sample of W-R stars is drawn from an observational program to study 
spectroscopically the cluster populations in M81 using the OSIRIS 
instrument at the GTC \citep[see e. g.][]{2013aMayya, 2013bMayya, 2014Mayya, 
2016Arellano}. The targets for this program were mainly Compact Stellar 
Clusters (CSCs) associated to recent star-forming sites catalogued by 
\cite{2010Santiago}. The entire dataset used for this work consists of 3 different 
observing runs, the first two using long-slits and the last one using MOS.
These observations were sensitive enough to register continuum and/or 
emission-line spectra of individual massive stars at the distance of M81.
This realization led us to carry out a comprehensive search for the W-R bumps 
not only at the position of the targeted clusters, but all along our reduced 
2-D spectral images of the long-slit. This search resulted in the discovery of 5 
objects along the 9 long-slits from the 2 runs in 2010 and 2012.

\begin{center}
\begin{figure*}
\includegraphics[width=12.7truecm]{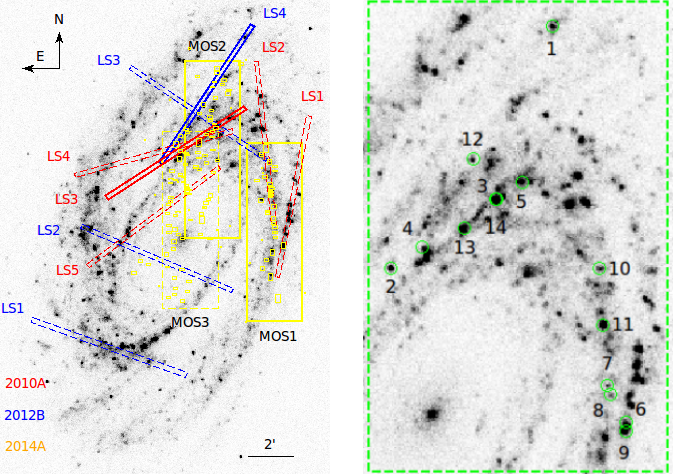}
\caption{\label{fig:galex_image} ({\it Left}) GALEX near-UV image of M81
showing the location of long-slits and MOS for the three different 
observing runs with GTC/OSIRIS. The long-slits are shown in red for 2010A 
and blue for 2012B, and the 3 different MOS fields are shown by yellow 
rectangles, each enclosing the individual slitlet positions. For the sake 
of clarity, slit-widths are amplified by a factor of 10 from the original 
width of $\sim$1.2\arcsec. The slits or MOS fields where the W-Rs are detected 
are shown by solid lines (2010-LS3; 2012B-LS4, 2014A-MOS1 and 2014A-MOS2). 
The orientation and a scale of 2\arcmin\ are also shown. 
({\it Right}) The spatial locations of the 14 detected W-R stars are 
indicated by green circles with their respective identification numbers.}
\end{figure*}
\end{center}

After observing an association of W-R detections with bubble morphologies, the 
slitlets for two of the 3 MOS pointings during the 2014 run was chosen to pass 
through CSCs surrounded by \ha\ bubbles/shells as seen in the F606W image 
of HST. Search for W-R features in the resulting spectra led 
to the discovery of 9 additional W-R objects. Thus, we have a combined sample 
of 14 objects with W-R features. Five of these objects coincided with the central 
position of CSCs in the sample of \cite{2010Santiago}, with the remaining objects 
appearing in the peripheral parts of the compact clusters or in open clusters. 
Figure~\ref{fig:galex_image} shows the positions of all the long-slits and the MOS 
fields on a GALEX\footnote{http://galex.stsci.edu/GR6/} image and the 
14 locations where the spectra registered W-R features. A log of all observations is given in 
Table~\ref{tab:table1}.

\begin{table*}
\caption{\label{tab:table1} Log of the long-slit and MOS spectroscopic observations with GTC/OSIRIS in M81.}
\begin{tabular}{llcrllccclr}
\hline
Run/Mode & PI  &Date &  PA   & SW   & Exp.Time  & AM    & Seeing &  Night & Std  & W-R\#\\
     (1)   &(2)  &(3)  &  (4)  &  (5)   &   (6)     & (7)   &   (8)  &   (9)  & (10) & (11)  \\
\hline 
2010A-LS1&D. Rosa-Gonz\'alez&2010-04-05&169.22&1.00&$3\times900$ &1.31&0.80&G&Feige34&--  \\
2010A-LS2&D. Rosa-Gonz\'alez&2010-04-05&  6.24&1.00&$3\times900$ &1.33&0.80&G&Feige34&--  \\
2010A-LS3&D. Rosa-Gonz\'alez&2010-04-05&123.23&1.00&$3\times900$ &1.41&0.80&G&Feige34&2--5\\
2010A-LS4&D. Rosa-Gonz\'alez&2010-04-05&105.20&1.00&$3\times900$ &1.56&0.80&G&Feige34&--  \\
2010A-LS5&D. Rosa-Gonz\'alez&2010-04-06&127.20&1.00&$3\times900$ &1.43&0.80&G&Feige34&--  \\
2012B-LS1&Y. D. Mayya&2013-01-12       &250.50&1.23&$3\times1500$&1.31&0.79&D&Feige34& -- \\
2012B-LS2&Y. D. Mayya&2013-01-12       &247.00&1.23&$3\times1500$&1.40&0.97&D&Feige34& -- \\
2012B-LS3&Y. D. Mayya&2013-01-12       & 56.10&1.23&$3\times1500$&1.32&1.20&D&Feige34& -- \\
2012B-LS4&Y. D. Mayya&2013-01-13       &146.58&1.23&$3\times1500$&1.35&0.95&D&Feige34&1   \\
2014A-MOS1&Y. D. Mayya&2014-04-03&0.00&1.20      &$3\times1308$&1.31&0.90&D&Ross 640& 6--11 \\
2014A-MOS2&Y. D. Mayya&2014-03-23&0.00&1.20      &$3\times1308$&1.35&1.00&D&Ross 640&12--14 \\
2014A-MOS3&Y. D. Mayya&2014-04-03&0.00&1.20      &$3\times1308$&1.34&0.80&D&Ross 640&  --  \\
\hline
\end{tabular}\\
Brief explanation of columns:
(1) Observing run (YYYYx-MOD\#, where YYYY is the year, x=Semester (A or B), MOD=Observing mode 
(LS=long-slit or MOS) and \# = Observing block number;
(2) Principal investigator; (3) Observational date (year-month-date); 
(4) Position angle ($^\circ$) of the slit as measured on the astrometrised image; 
(5) Slit-width (\arcsec); 
(6) Exposure time (number of exposures $\times$ integration time in seconds); 
(7) Mean airmass of the 3 integrations; 
(8) Seeing (\arcsec);
(9) Night (G = grey or D = dark), clear skies (cirrus reported only for 2010A-LS5); 
(10) Standard star name; 
(11) Detected W-R ID number.\\
\end{table*}

Both the long-slit and MOS observations were carried out using R1000B grism
with a slit-width of $\sim$1.2\arcsec, covering a spectral range from $\sim$3700 
to 7500~\AA, at a spectral resolution of $\sim$7~\AA. The observations 
were carried out with a CCD binning of $2\times2$ resulting in 
a spatial scale of 0.254\arcsec\,pixel$^{-1}$ (horizontal axis) and spectral
sampling of $\sim$2\,\AA\,pixel$^{-1}$ (vertical axis). The long-slits were 
7.4$^\prime$ long, whereas each MOS observation included $\sim$30--40 slitlets 
of varying lengths over a field of view of $\sim$7.4$^\prime\times2^\prime$. 
These slitlets included object-free regions for subtraction of the sky spectra, 
and additional $\sim$5--6 stars as fiducial points for astrometry. 

All the observations were carried out in the service mode, with the
total observing time split into blocks of $\sim$60--90~min. Each block
corresponded to a particular long-slit position or a MOS pointing, and consisted of
3 exposures of equal integration time, to facilitate later removal of 
cosmic ray hits. The observing runs involved a total of 12 observing blocks. 
Data for each block contained ancillary files that included standard stars, 
bias, flat-field and arc lamps. 
The sky was stable during all the observations where W-R bumps were detected.

\subsection{Spectroscopic reductions}

Reduction of spectroscopic data was carried out using the
package {\it GTCMOS}, a tailor-made IRAF-based pipeline developed by one of us 
(YDM)\footnote{http://www.inaoep.mx/$^\sim$ydm/gtcmos/gtcmos.html}
for the reduction of the GTC/MOS and long-slit spectra. 
The pipeline uses the standard spectroscopic tasks available in 
IRAF\footnote{IRAF is distributed by the National Optical Astronomy Observatory, 
which is operated by the Association of Universities for Research in Astronomy 
(AURA) under cooperative agreement with the National Science Foundation.}  
to carry out the reductions. As this is the first paper that uses this pipeline,
we will explain briefly the reduction procedure below.

The OSIRIS instrument registers the spectra on 2 CCDs, with the dispersion axis
running vertically on a single CCD. As a first step of the reduction procedure, 
the pipeline creates a single spectral image of $2110\times2051$ pixels, by tiling 
the two individual CCD images, using an IRAF script provided by the OSIRIS team. 
The geometrical distortions of the images were corrected in this step. 
A master bias image is created for each observing block by combining all tiled
bias frames using median algorithm, which is then subtracted from all images of 
that block. The multiple images of the target objects (3 in our case) within a 
single observing block were then combined using the median algorithm, in the 
process cleaning the image of cosmic ray events.
Each observing block contained multiple exposures of arc-lamp spectra of 
Ne, Hg and Ar for both long-slit and MOS observations. The arc spectra in the
two CCDs were tiled into a single spectral image. The tiled images of all available
arc spectra were summed to get a single arc image, which is then used for 
wavelength calibration.
GTCMOS tasks {\it omstart}, {\it omcombine} and {\it omidentify}
accomplish the above described functions. 

The slit image in the spatial direction has significant curvature. 
The task {\it omidentify} straightens this curvature by analysing one arc-lamp
spectrum for every section of 20~pixels in the long-slit mode,
or at the center of every slitlet in the MOS mode.
Line identification and centering were carried out by an automatic routine that 
uses $\sim$10 bright unblended lines in the useful spectral range for each grism.
Independent dispersion solutions are obtained for every MOS slitlet (or for image 
section of 20~pixels in the long-slit mode) with the IRAF task {\it identify} 
using a {\it spline3} function of order 2. Solutions that resulted in an rms 
error of $>$0.5\,\AA\ were improved by manually examining the fitted residuals.
Final rms errors in all cases remained better than 0.5\,\AA, with the majority
having rms errors $<$0.2\,\AA. The best-fit solution for each image section is used 
to create a wavelength calibrated 2-D image, where the spectral axis was linearly
re-sampled to have a common dispersion of 2.1\,\AA\,pixel$^{-1}$, the mean
resolution for the grism R1000B.
Tilts/curvatures within a MOS slitlet or in a 20-pixel section of a long-slit
are corrected by linearly shifting the dispersion-corrected spectra 
so as to force the centroid of the [OI]$\lambda$5577 sky line in every spectrum 
in that section to its rest wavelength. 
The procedure followed by the pipeline ensured that the skylines
are perfectly horizontal in the 2-D spectral image for both long-slit and MOS modes. 
The GTCMOS task {\it omreduce} accomplishes these functions.

The basic reduction of the standard star spectra followed a procedure similar to that
described above. 
All independent standard star exposures within an observing block were reduced 
individually. The spectra were extracted and the IRAF task {\it standard} was 
used to obtain the sensitivity tables between the flux and count rate at all 
available spectral bands of the catalogued standard star. The sensitivity
tables for all exposures within an observing block were averaged at every
wavelength band and fitted with a spline3 function of order 6 using the IRAF task 
{\it sensfunc} to obtain a mean sensitivity function. The extinction curve
for the observatory, along with the observed airmass, were used for the 
purpose of atmospheric extinction correction. 

M81 is a large galaxy even for the relatively large field of view of the
OSIRIS instrument, which implies that the galaxy disk has contribution even in the
designated sky slitlets in the MOS mode or the apparently object-free pixels
of the long-slit. We examined the HST images at the positions of the slits to 
search for pixels that are devoid of point sources. Spectra of such sections
were used to obtain a sky + disk spectra, 
which were then subtracted from each object spectrum.
In regions with varying disk contribution, care was taken to subtract the 
disk spectrum spatially closest to the object. 

We found that the flat-fields in the blue part of the spectrum do not have 
the signal-to-noise ratio suitable for the correction of the relative 
efficiencies of pixels, even when the red part of the spectrum is saturated.
Thus, in order to avoid degrading the blue part of the target spectra, we carried out the
reductions without applying corrections for flat-fielding and illumination
variations along the long-slit, or from one slitlet to another in the MOS mode.
The flat-field errors are expected to be less than a percent, whereas
the residual error due to the non-correction for illumination variations
is estimated to be at the most 2\% percent, as judged by the relative 
intensities of the skyline along the spatial direction.

\subsection{Detection of W-R features and extraction of spectra}

We visually examined the wavelength-calibrated and sky-subtracted 2-D spectral image
from all the runs for the presence of blue and/or red bumps. This procedure
resulted in the identification of 14 regions with W-R features.
Continuum is detected at the location of all the 14 W-R features, with the
continuum having a higher spatial extent than that of the W-R feature in some cases.
Typically the width of the blue bump in the spatial direction was measured to
be of the seeing size (4-5~pixels). On the other hand, the nebula as traced
by the \hb\ line extended much more in the spatial direction.

We followed a similar procedure for the extraction of 1-D spectrum from our 
long-slit and MOS 2-D spectral images. The IRAF task {\it apall} was used for 
this purpose. We tuned the {\it apall} parameters so as to maximize the strength 
of W-R bumps in the extracted 1-D spectrum. This is achieved by centring
the extraction window on the peak of the blue bump and the 
spectra were traced around this pixel. In a few cases (3) that had a weak
continuum, we used a bright nearby continuum object as a reference for tracing,
whereas in a few others that had a broad continuum, we restricted
the trace width to 5~pixels. A residual sky is subtracted by
choosing a local sky region interactively.

\begin{figure*}
\begin{centering}
\includegraphics[width=8.3truecm]{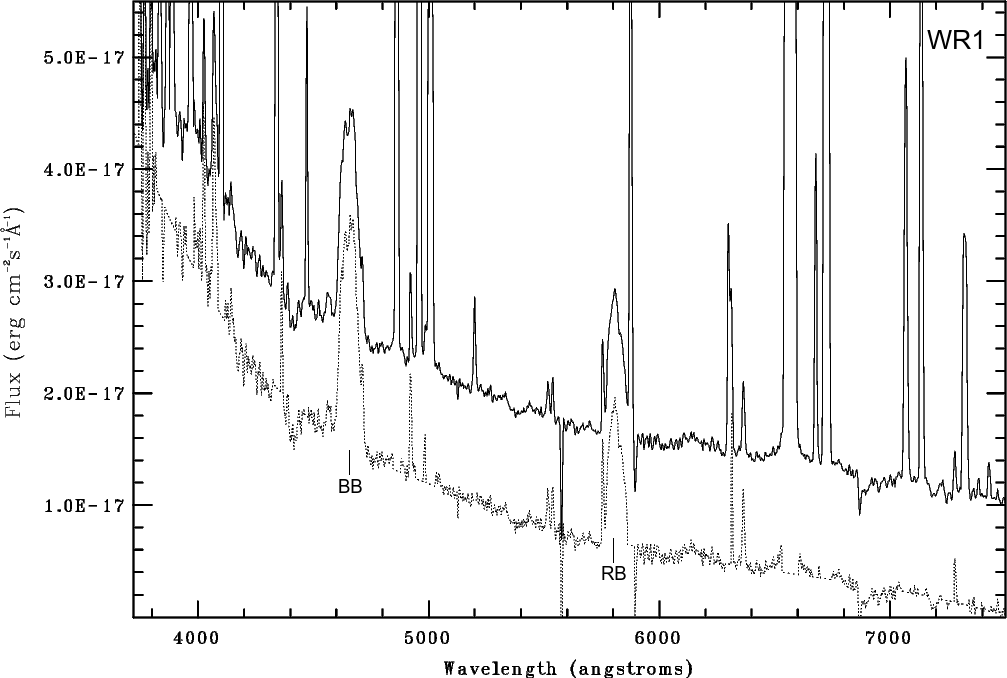}
\includegraphics[width=8.3truecm]{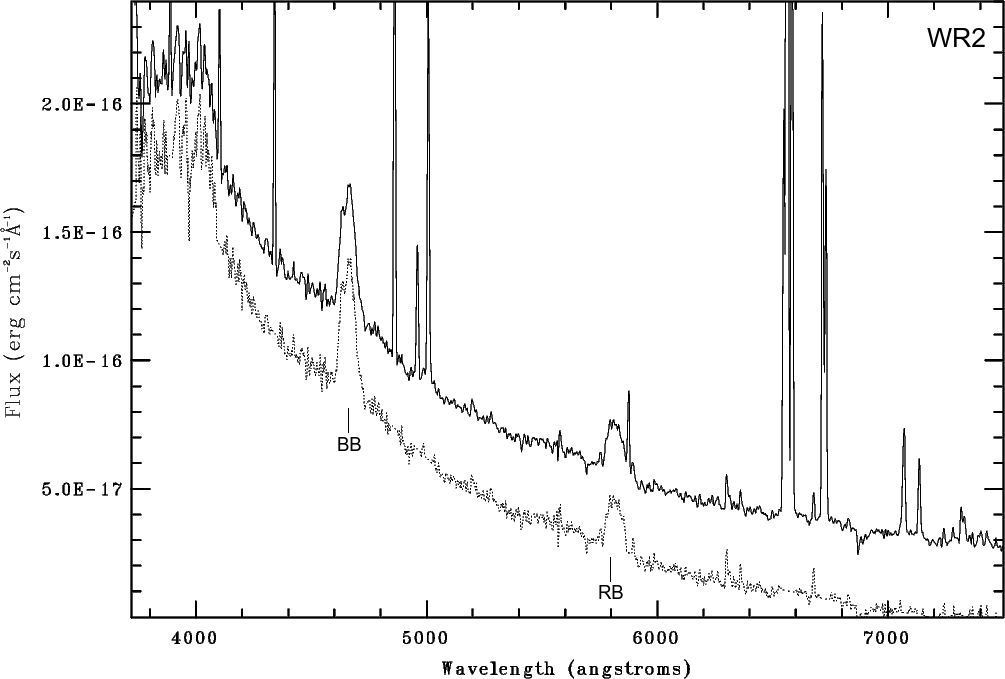}
\includegraphics[width=8.3truecm]{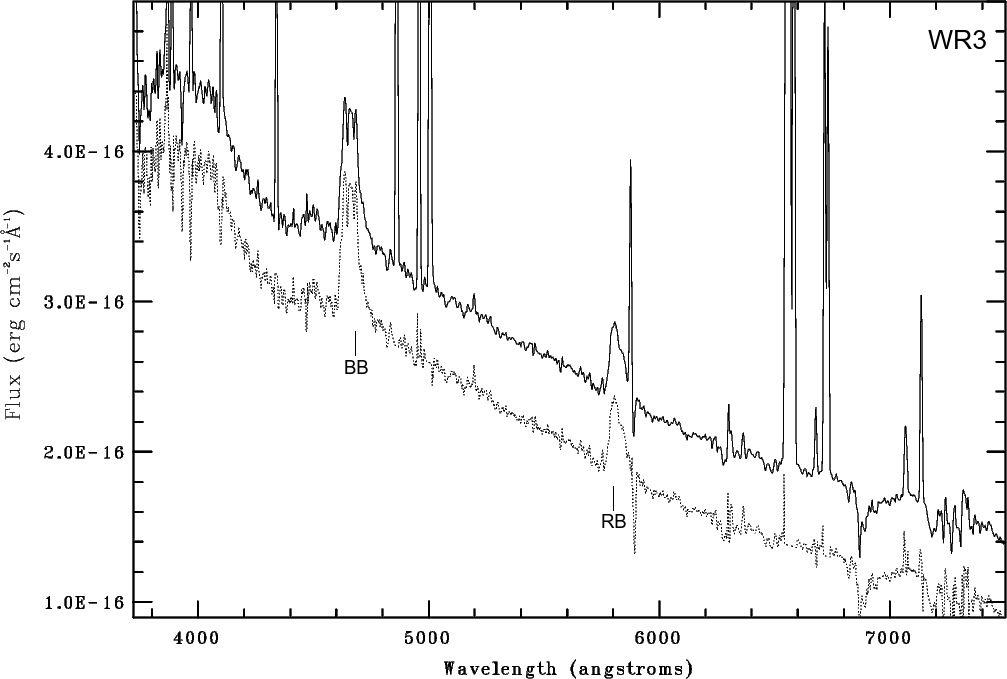}
\includegraphics[width=8.3truecm]{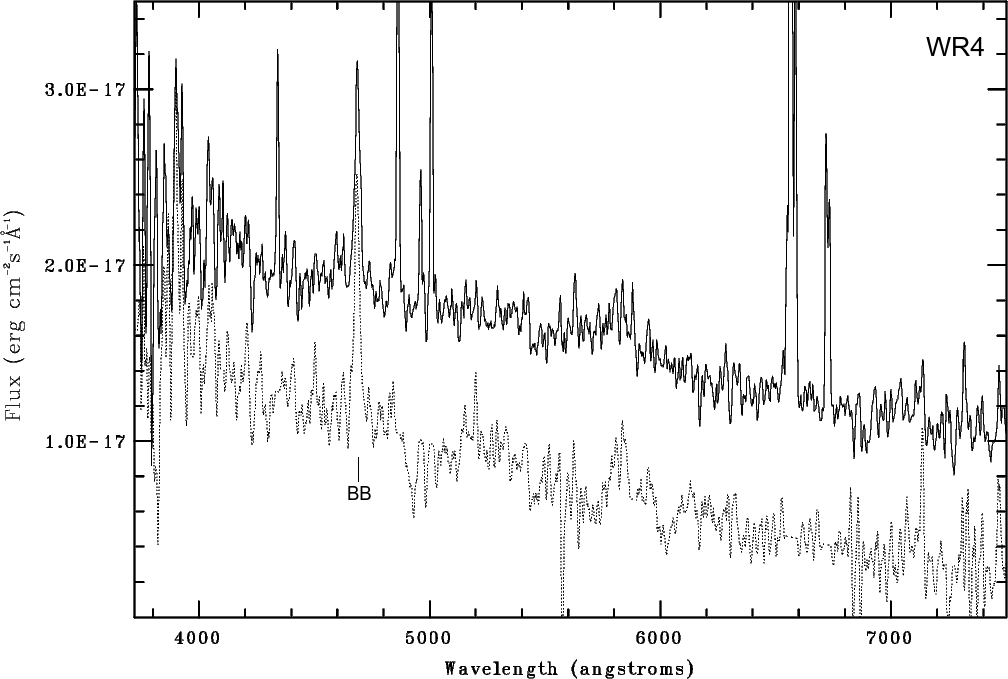}
\includegraphics[width=8.3truecm]{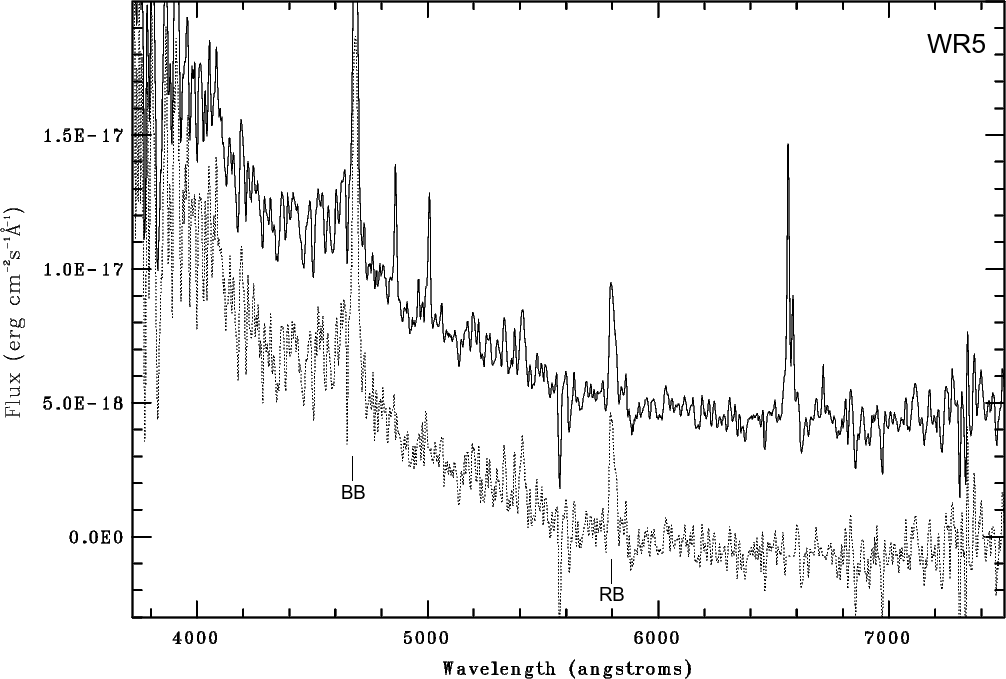}
\includegraphics[width=8.3truecm]{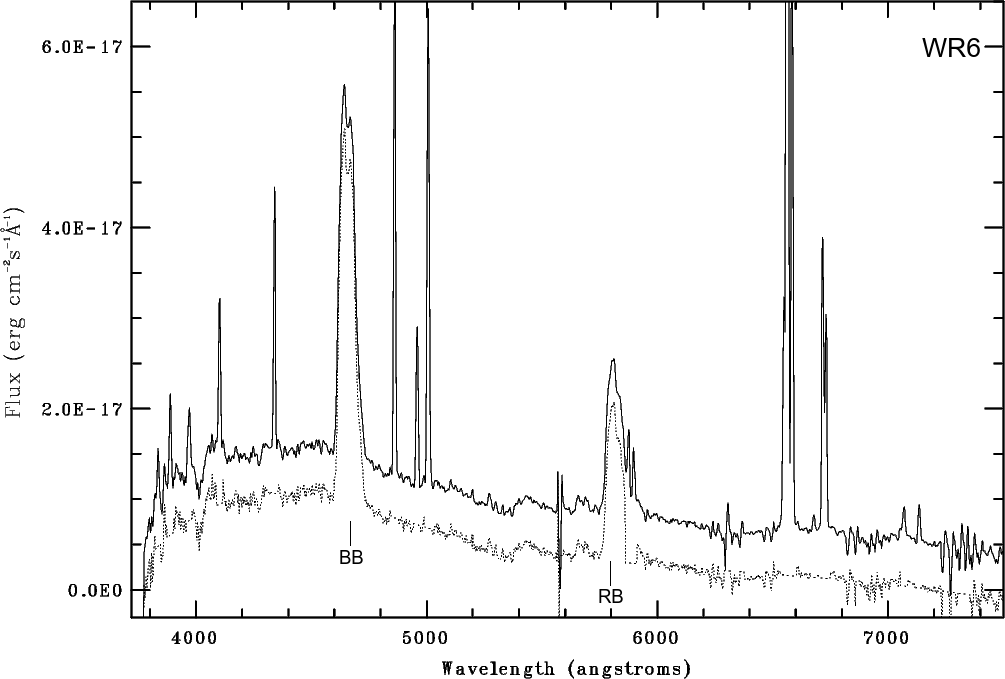}
\includegraphics[width=8.3truecm]{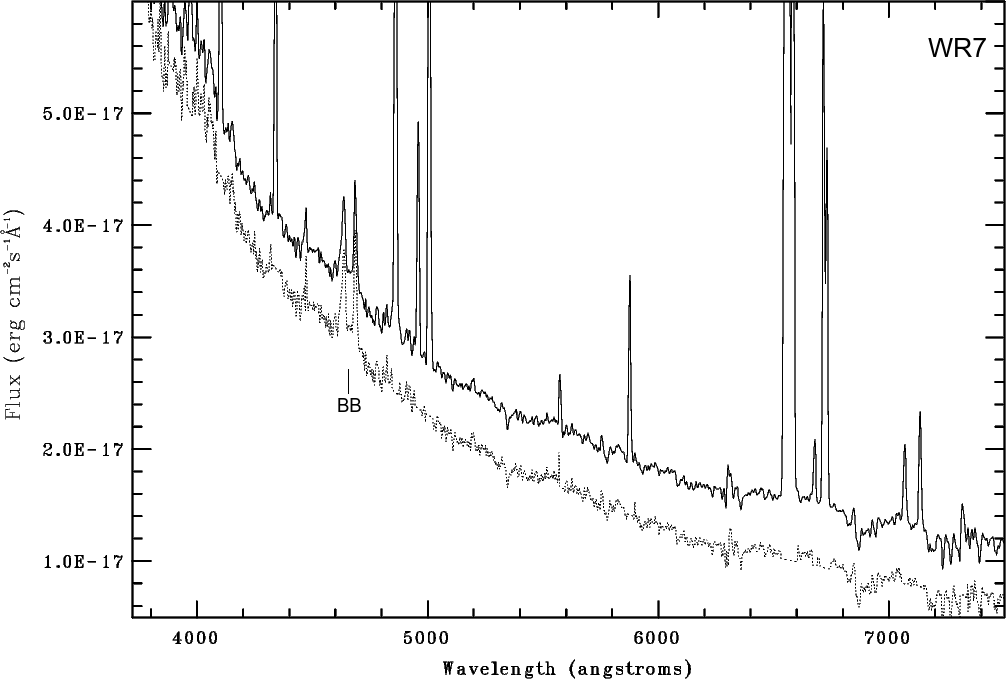}
\includegraphics[width=8.3truecm]{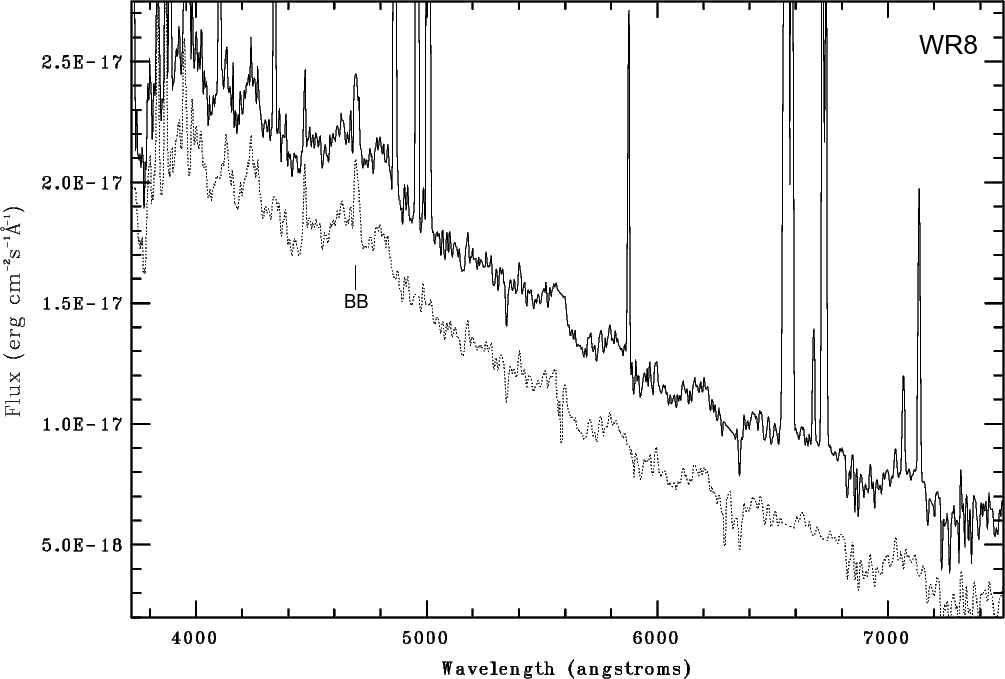}
\par\end{centering}
\caption{\label{fig:spectra1} Extracted spectrum of each of the 14 locations containing W-R features
(above) is plotted along with the corresponding nebular-free spectrum (below) in each panel.
The latter spectra are shifted downwards by appropriate amounts for the sake of clarity.
The top right corner of each panel contains the W-R identification.
The blue and red bumps, when present, are marked by letters BB and RB, respectively.
}
\end{figure*}

All extracted spectra are rich in nebular emission lines,
originating in an ionized nebula around the W-R stars. Bright nebular lines in these spectra were used to determine the radial
velocity for each spectrum, which are given in the last column of Table~\ref{tab:table2}. All 
spectra were brought to the rest-frame wavelength using these measured velocities.
The Doppler-corrected spectra for all the 14 W-R stars are shown in Figure~\ref{fig:spectra1}.
The blue and red bumps can be easily noticed around 4650\,\AA\ and 5810\,\AA.

It is desirable to have W-R spectra free of nebular lines for an easy identification
of all features responsible for W-R stars.
Our slits registered nebular spectra from regions surrounding the W-R 
sources. These spectra are Doppler-corrected using the velocities measured in these spectra,
and are scaled using the fluxes of the \hb\ line, before using them for subtracting the nebular contamination in the W-R spectrum. 
The residual fluxes of the nebular lines in the resulting pure W-R spectrum
are around 20\% of the original flux, which limits the detection of any W-R line 
coincident with a nebular 
line (e.g. \ha, \hb, \civheir\ etc.). In order to avoid any misinterpretation of
the residuals, we replaced the part of the spectrum containing a nebular line 
with continuum fitted to the original W-R spectrum. 
The adopted technique allows us to easily identify W-R lines that are not 
coincident with the lines in the surrounding nebula. The resulting nebular-free spectra are plotted below the observed spectra in Figure~\ref{fig:spectra1}. Analysis of the W-R features in these spectra will be presented in \S3.

\subsection{Localization of W-R stars using the HST images}

\begin{figure*}
\begin{centering}
\includegraphics[width=8.3truecm]{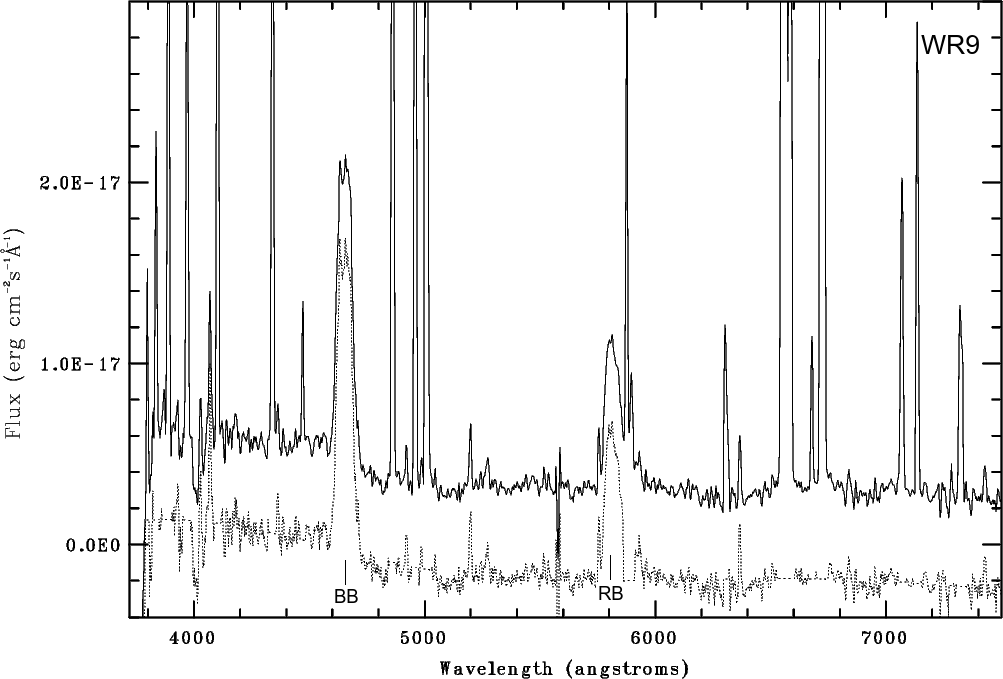}
\includegraphics[width=8.3truecm]{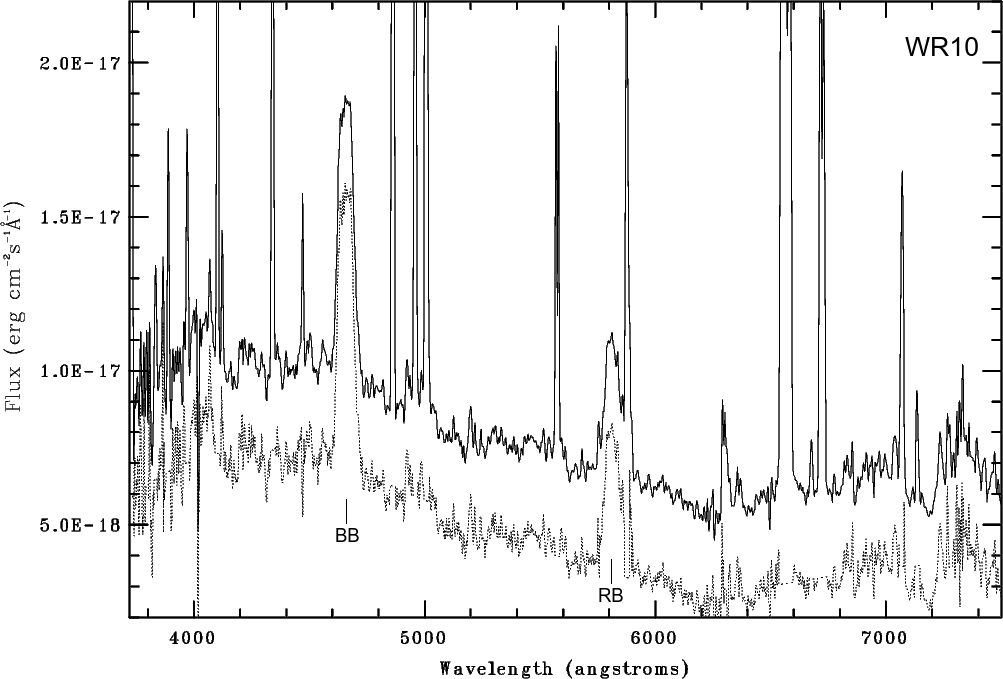}
\includegraphics[width=8.3truecm]{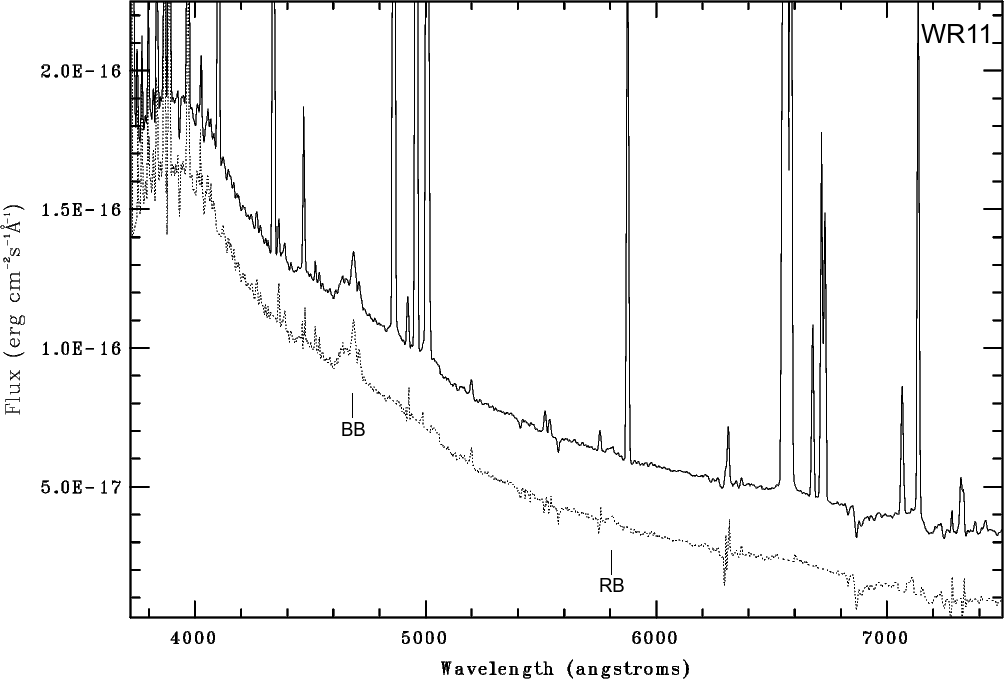}
\includegraphics[width=8.3truecm]{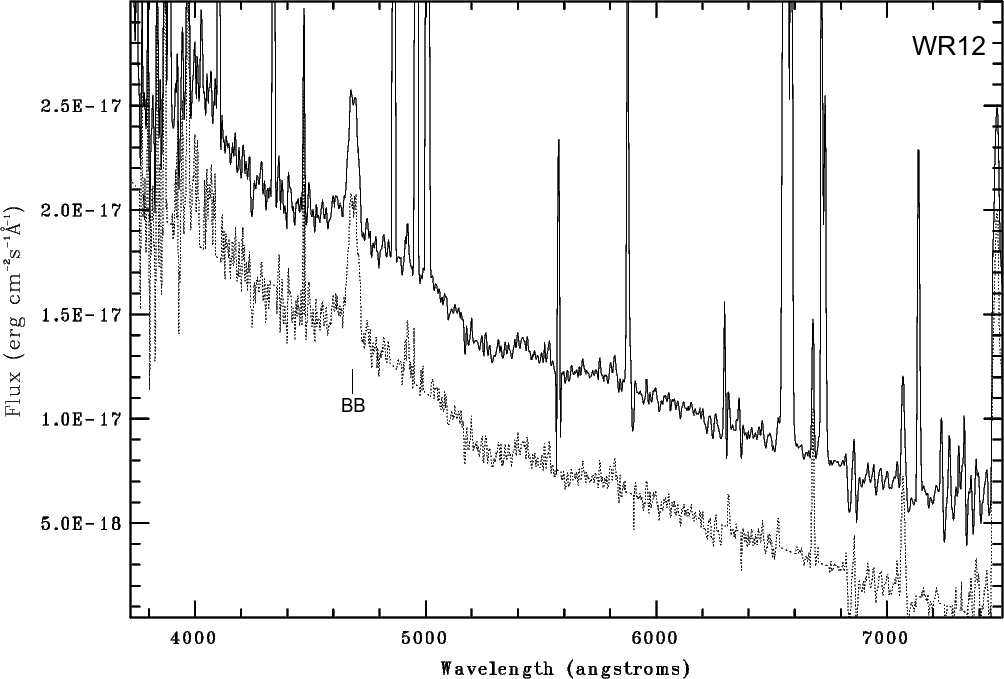}
\includegraphics[width=8.3truecm]{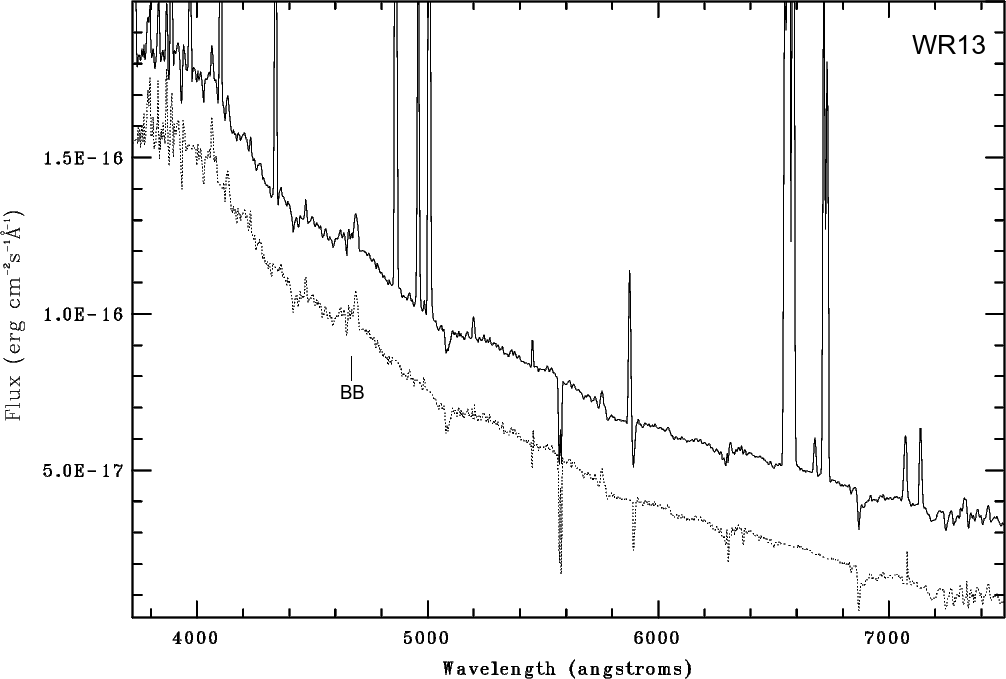}
\includegraphics[width=8.3truecm]{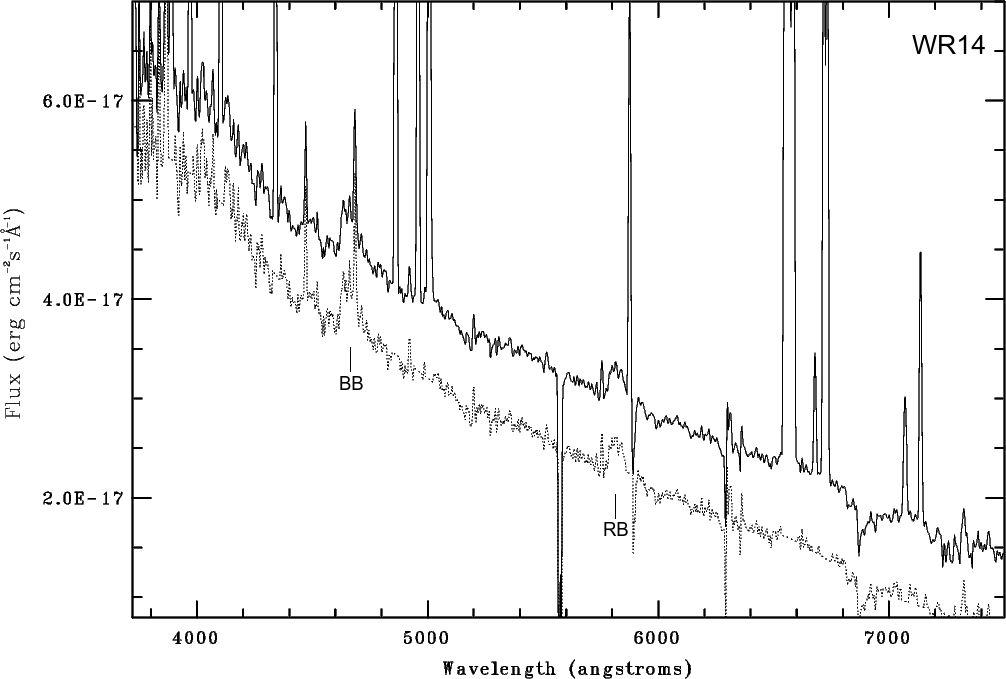}
\par\end{centering}
{\bf Figure~2} continued.
\end{figure*}

The physical scale of our slit-width ($\sim$20~pc) is large enough to include more 
than one star, and hence the W-R features seen in our spectra may not be 
necessarily originating in single stars. We hence use
the 10 times better spatial resolution offered by the HST/ACS images 
in an attempt to identify the object(s) responsible for W-R features.
The HST database has images in the B (F435W), V (F606W) and 
I (F814W) (PI: Andreas Zesas) bands for M81, all using the ACS. 
The F606W filter intercepts the \ha\ line, which allows us 
to trace the nebular emission, if any, at the HST resolution.

\begin{figure*}
\begin{centering}
\includegraphics[width=5.5truecm]{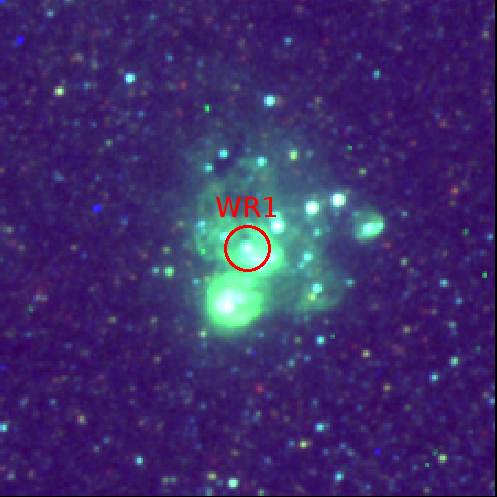} 
\includegraphics[width=5.5truecm]{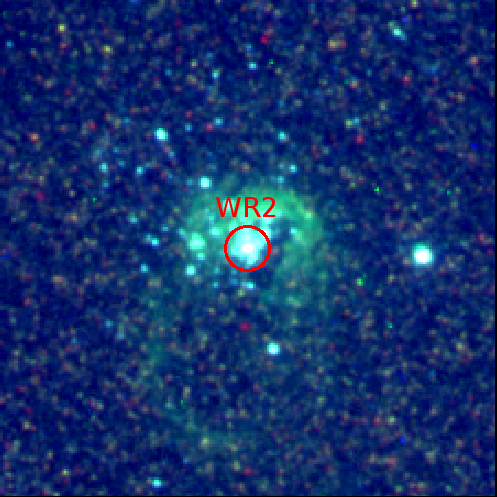} 
\includegraphics[width=5.5truecm]{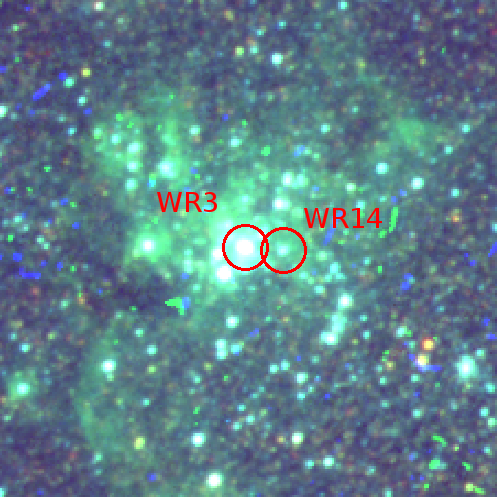}\\
\includegraphics[width=2.71truecm]{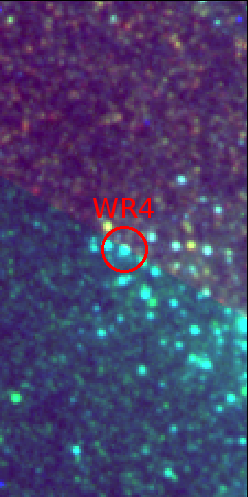} 
\includegraphics[width=2.71truecm]{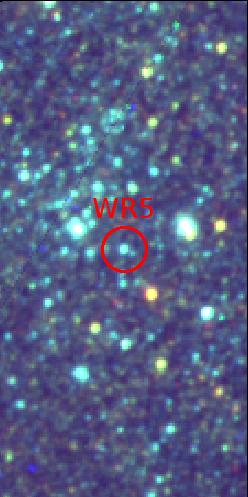} 
\includegraphics[width=5.5truecm]{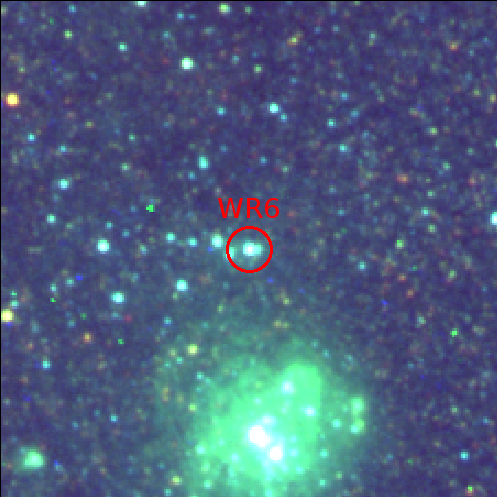} 
\includegraphics[width=5.5truecm]{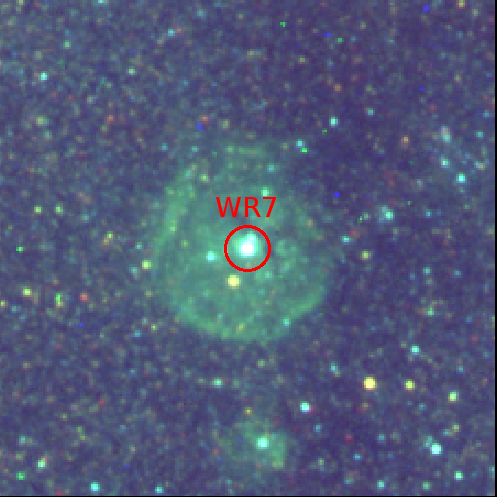} \\
\includegraphics[width=5.5truecm]{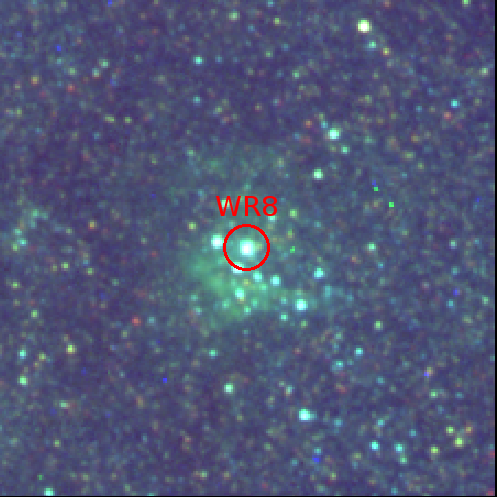} 
\includegraphics[width=5.5truecm]{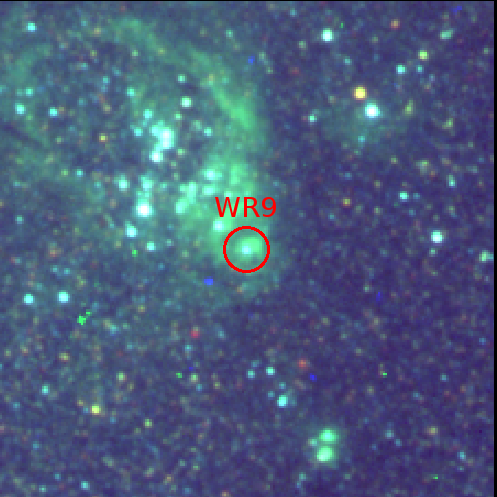} 
\includegraphics[width=5.5truecm]{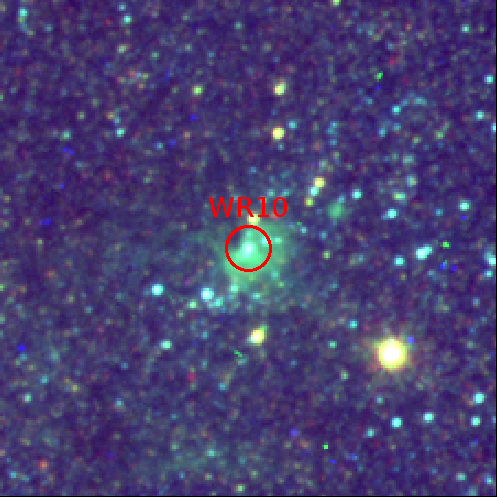} \\
\includegraphics[width=5.5truecm]{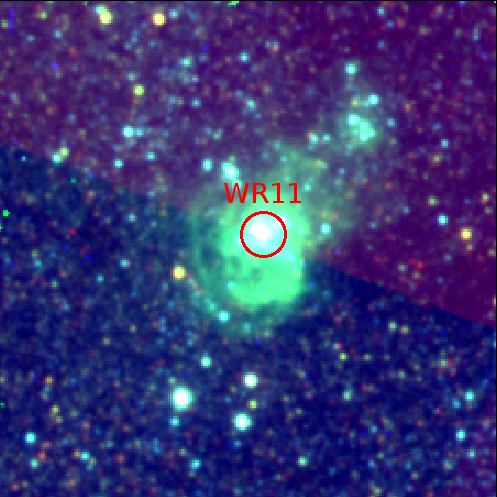} 
\includegraphics[width=5.5truecm]{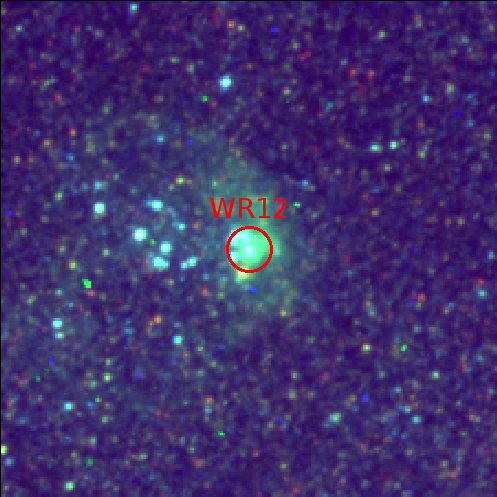} 
\includegraphics[width=5.5truecm]{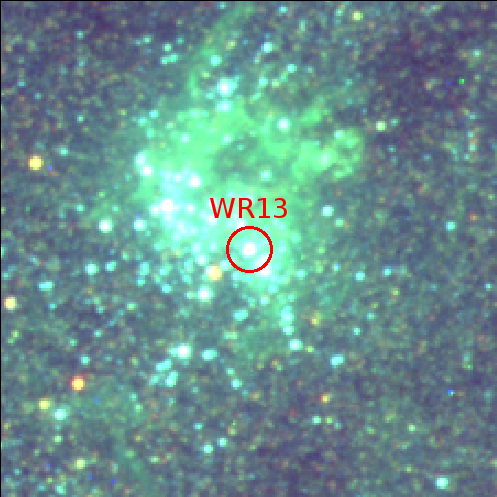}\\
\par\end{centering}
\caption{\label{fig:nebula} RGB images using the 
HST/ACS F814W (R), F606W (G) and F435W (B) bands showing the large-scale morphology 
of ionized gas and stars around the W-R star locations. The sections are of physical size
200$\times$200~pc in all cases except in WR4 and WR5 where the sizes correspond
to 100$\times$200~pc. The W-R locations are shown by red circles. The ionized gas is seen as
green emission in these images, which is due to the \ha\ line entering the
F606W filter. All images are oriented north at the top and east to the left.
}
\end{figure*}

The following procedure was adopted for locating the W-Rs on the HST images.
All the GTC observing blocks contained two acquisition images in the SDSS $r$-band; 
one of the slit, and the other of the entire field of view of
$\sim$60~sec and 10~sec durations, respectively. We astrometrised the latter
image with respect to the HST image using $\sim$20 field stars, giving us an astrometric
accuracy of $\sim$0.1\arcsec\ rms. We then superposed the long-slit/MOS slitlets 
on an RGB colour image formed from 
the F814W (R), F606W (G) and F435W (B) images. On this image, we searched for the
W-R star around the position calculated in the acquisition image within an error
box of 0.1\arcsec$\times slit-width$. 
In majority of the cases, the brightest object is the
closest to the expected position. Often, this object occupies the centre of the 
nebula, and the nebular size measured using the \ha\ line in the spectrum 
is in agreement with that measured on the F606W image. In 3 cases, there were 2--3 
candidates within the search box. In such cases, we 
selected the one closest to the expected position and/or the brightest 
object within the search box. The positions of the identified W-R stars are 
marked in Figure~\ref{fig:nebula}, where the diameter of the circle corresponds
to 1\arcsec, which is the typical slit-width in our observations.

\begin{table*}
\begin{center}
\caption{\label{tab:table2}Sample of W-R stars in M81.}
\begin{tabular}{llllrrlllrrrr}
\hline
ID &  R.A. & DEC   & $V$  & $B-V$  & $B-I$ & $M_V$ & Complex & Cluster & D &  Size$_{\rm neb}$ & V$_{\rm helio}$\\
(1)&    (2)    &(3)        & (4)   & (5)     & (6)     & (7)   & (8) & (9) & (10) & (11) & (12)\\
\hline
WR1& 148.75693 & 69.215993 & 21.29 &   0.54  &   0.24  & $-6.51$ &Munch 18 &F04B16353 &1.0& 100 &38\\
WR2& 148.93458 & 69.122063 & 20.54 & $-0.14$ & $-0.26$ & $-7.27$ &R06B06945&R06B06945 &0.0& 100 &50\\
WR3& 148.81938 & 69.148725 & 19.53 &   0.06  &   0.09  & $-8.27$ &kauf152  &R03B16992 &0.0& 250 &137\\
WR4& 148.89954 & 69.130043 & 22.64 &   0.02  & $-0.66$ & $-5.16$ & --      & --       &-- & --  &58\\
WR5& 148.79077 & 69.155475 & 23.07 & $-0.19$ & $-0.41$ & $-4.73$ & --      &R03B08603 &1.5& --  &191\\
WR6& 148.67759 & 69.061852 & 21.56 & $-0.08$ & $-0.21$ & $-6.24$ &kauf125  &R04B15666 &5.0& 20  &9\\
WR7& 148.69802 & 69.076172 & 20.46 &   0.18  &   0.37  & $-7.34$ &kauf127  &R04B07520 &0.0& 80  &43\\
WR8& 148.69482 & 69.072919 & 21.45 &   0.06  &   0.10  & $-6.35$ & --      &R04B08657 &0.0& 80  &25\\
WR9& 148.67736 & 69.058316 & 22.37 &   0.40  &   0.20  & $-5.43$ &kauf125  &--        &-- & 100 &$-14$\\
WR10&148.70591 & 69.121952 & 21.90 &   0.54  &   0.59  & $-5.90$ &kauf135  &R02B18517 &0.0& 80  &110\\
WR11&148.70268 & 69.099903 & 19.30 &   0.17  & $-0.12$ & $-8.50$ &kauf128  &R04B01345 &3.0& 80  &90\\
WR12&148.84451 & 69.164518 & 21.78 &   0.35  &   0.20  & $-6.02$ & --      &--        &-- & 100 &86\\
WR13&148.85411 & 69.137450 & 20.22 &   0.14  &   0.20  & $-7.58$ &kauf159  &R03B25826 &3.0& 250 &104\\
WR14&148.81870 & 69.148704 & 22.17 &   0.19  &   0.11  & $-5.63$ &kauf152  &R03B16992 &2.0& 250 &192\\

\hline
\end{tabular}\\
Brief explanation of columns:
(1) Name of the W-R star adopted in this study;
(2--3) Right Ascension and declination in the FK5 system on the astrometrised 
HST image, where the M81 nucleus is located at RA=148.88889$^\circ$ (9:55:33.333) 
and Dec=69.065333$^\circ$ (+69:03:55.20)
(4-6) Apparent magnitude and colours in F435W (B), F606W (V) and F814W (I) bands;
(7) Absolute magnitude in F606W band ($M_V$) using a distance modulus of 27.80~mag, and Galactic extinction A$_{\rm V}=$0.20~mag \citep{2011Schlafly};
(8) Star-forming complex from \cite{2006Perez} embedding the W-R star;
(9) Compact stellar cluster from \cite{2010Santiago} within 100~pc (5.7\arcsec) 
that is closest to the W-R;
(10) Distance to the compact stellar cluster in \arcsec;
(11) Size in parsec of the nebulosity around the W-R star as measured on the F606W image;
(12) Heliocentric radial velocity in km~s$^{-1}$ of the nebular lines in the extracted W-R spectrum.\\

\end{center}
\end{table*}

\subsection{Large-scale morphology around the W-R stars}

We now examine the large-scale environment around the identified W-R locations. 
For this purpose, we use the RGB images generated using the HST/ACS F814W (R), 
F606W (G) and F435W (B) bands (Figure~\ref{fig:nebula}). In particular, we 
use the F606W image which traces the nebular emission (seen as green colour 
in our RGB images). Large-scale ($\sim$100~pc) ionized gas is seen in all but 2 cases, 
with the exceptions being WR4 and WR5. Small-scale ($<$20~pc) ionized gas is present
however even in these two cases as can be inferred from the nebular emission lines
in their spectra.

All W-R stars in our sample are part of a star-forming complex, or an open cluster.
In column~8 of Table~\ref{tab:table2} we list the name, when available, of the 
star-forming complex containing the W-R star from \cite{2006Perez}. The presence
of a cluster can be inferred in the HST image even in the 3 W-R locations without an 
association to a named complex. If the W-R location is part
of the CSC or close to it, we list the name of the CSC from \cite{2010Santiago}
in column~9. Column 10 contains the distance of the W-R star from the nearest 
CSC. In 5 cases, the W-R star occupies the core of the compact cluster.
Two of our W-R locations (WR3 and WR14) are part of a same complex, whereas one (WR6) is 
clearly outside a CSC, with the nearest CSC at a distance of about 90~pc. 
This is most likely a runaway W-R star, which are found at distances
as far as 120~pc from the parent clusters \citep[e.g. runaway O star in the dense 
LMC cluster R136 in 30~Dor complex;][]{2010Evans}.

The ionized gas has a symmetric bubble morphology in one case (WR7), whereas 
in rest of the cases, a shell-like structure extends 
on one side of the cluster. The sizes of the bubbles/shells range from 
$\sim$50--250~pc (see column 11). 

\subsection{Photometry of W-R objects}

Having located the W-R objects on the HST images, we carried out 
aperture photometry of these objects using IRAF/phot. An aperture radius
of 0.2\arcsec\ was used, and an aperture correction factor for an
infinite aperture was applied, along with their zeropoints 
(F435W: 0.03, 25.579; F606W: 0.03, 25.00; F814W: 0.03, 25.501).

\begin{figure}
\begin{centering}
\includegraphics[width=8.0truecm]{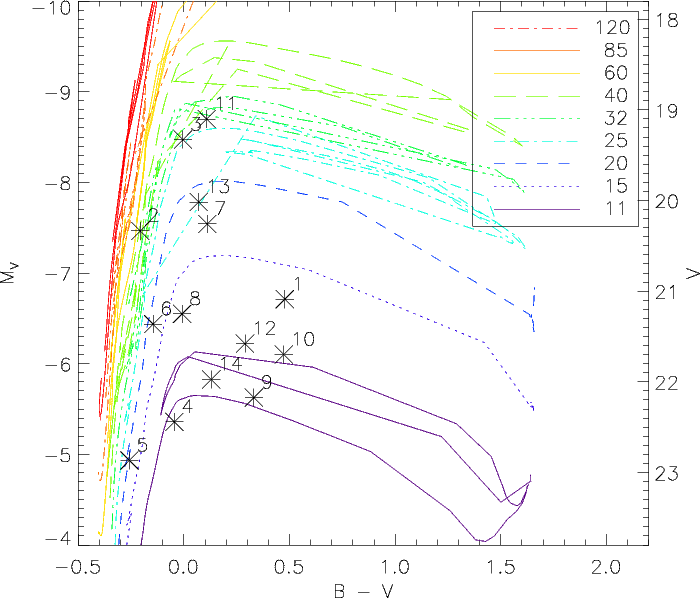}
\par\end{centering}
\caption{\label{fig:cmd}Colour-Magnitude Diagram (CMD) for the 14 W-R stars 
in M81 (asterix), each identified by its number. The left axis shows the
absolute magnitude, whereas the right axis shows the observed magnitude, both
in the HST F606W (V) band. Evolutionary tracks at solar metallicity from
Geneva models \citep[][standard mass-loss and no rotation]{2012Ekstrom} are shown 
for masses between 11 (bottom most) and 120~\msol\ (top most), indicated with 
different line types and colours at the top-right corner. 
}
\end{figure}

A colour-magnitude diagram using these data is shown in Figure~\ref{fig:cmd}, 
where we compare their positions with evolutionary tracks for massive stars
at solar metallicity from Geneva models \citep[][standard mass-loss and no 
rotation]{2012Ekstrom}.  
The plotted models indicate that the magnitudes and colours of our objects 
are consistent with them being individual stars.  
The evolutionary tracks indicate masses of our 
W-R stars between $\sim$11 and 120~\msol. However, in these models W-R 
properties (hydrogen abundance $<40$\% and $T_{\rm eff}>10000$~K)
are generated only for masses $>32$~\msol. Inclusion of rotation in models,
or binarity can resolve this apparent problem regarding the minimum mass that
goes through the W-R phase. The observed range of magnitudes and colours are 
in the range of the values for individual W-R stars in the 
MW \citep[see Table~2 in][and references therein]{2007Crowther}.

\section{Analysis of W-R features}

The blue and red bumps of the W-R stars are made up of several broad
emission lines. It is customary to relate the blue bump to WN subtype,
being dominated by the \heiiwr\ and \niiiwr, whereas the red bump 
is associated to the \civwrr\ line from the WC subtype. 
The blue bump has a contribution from \ciiiwrb/\civwrb\ line as well. 
Additionally, nebular lines of \heiiwr, \civheib, \niineb\ and \civheir\ can 
be present superposed on the broad bumps.

WN and WC stars are further classified into subtypes WN2--9 and WC4--11
\citep{1968Smith, 1996Smith, 1998Crowther}. WN2--5 and WC4--6 are known 
as early (WNE and WCE), with the higher numbered subclasses referred to as 
late (WNL and WCL) types. Early types show lines of higher ionization 
state (\niv, \nv\ in WNEs; \civ\ in WCEs) as compared to the late types (\niii\ in WNL; \ciii\ in WCL). 
\citet{1994Smith} has extended the WN  classification in order to include
Of stars. These stars are given types WN10--WN11. Often, these stars show
hydrogen lines in emission and/or absorption.

For an accurate classification of subtypes
in each of WN and WC classes, it is important to determine the strength
of the diagnostic lines without contamination from other lines. 
Often the quality of the available spectra does not permit identification
of all diagnostic lines. In such cases, subtypes have been determined by 
making use of templates either obtained observationally 
\citep[e.g.][]{2006Crowther} or theoretically \citep[e.g.][]{2004Hamann}.

The nebular-subtracted spectra are analysed using template spectra of individual
W-R stars in order to determine (1) the number of W-R stars responsible for the 
observed strength of W-R features, and/or (2) the W-R subtype(s) of the contributing
star(s). We used the observed templates of LMC stars\footnote{The metallicity of
the observed regions in M81 resembles that of the Milky Way. However, as of now
templates are available only for LMC and SMC W-R subtypes.},
that are kindly provided to us by Paul Crowther \citep{2006Crowther}.
The template spectra were converted from luminosity to flux units by using 
the distance of M81. The resulting continuum level was found to be in general higher 
and bluer than the observed continuum of M81 W-R spectra. We hence added
a power-law continuum to the template so as to reproduce the observed 
continuum. For our spectra containing a red bump (8 cases), we used the WC template and 
scaled it by a factor \fwc\ so as to match the observed strength of the red bump. 
In 2 cases, we found the scaled template also reproduced very well the observed
strengths of the blue bump. These are pure WCE stars.
In the remaining 6 cases, we found that there is a systematic residual on 
the blue edge of the observed blue bump. In these cases, we added a WNL or WNE 
template spectrum and scaled it by another factor \fwn\ so as to reproduce 
the observed blue bump strength. In rest of the cases with only a blue bump, 
a scaled WN template is found to reproduce the observed bump profiles very well.
Results of the classification are summarized in Table~\ref{tab:table3}.

\begin{figure*}
\begin{centering}
\includegraphics[width=8.5truecm]{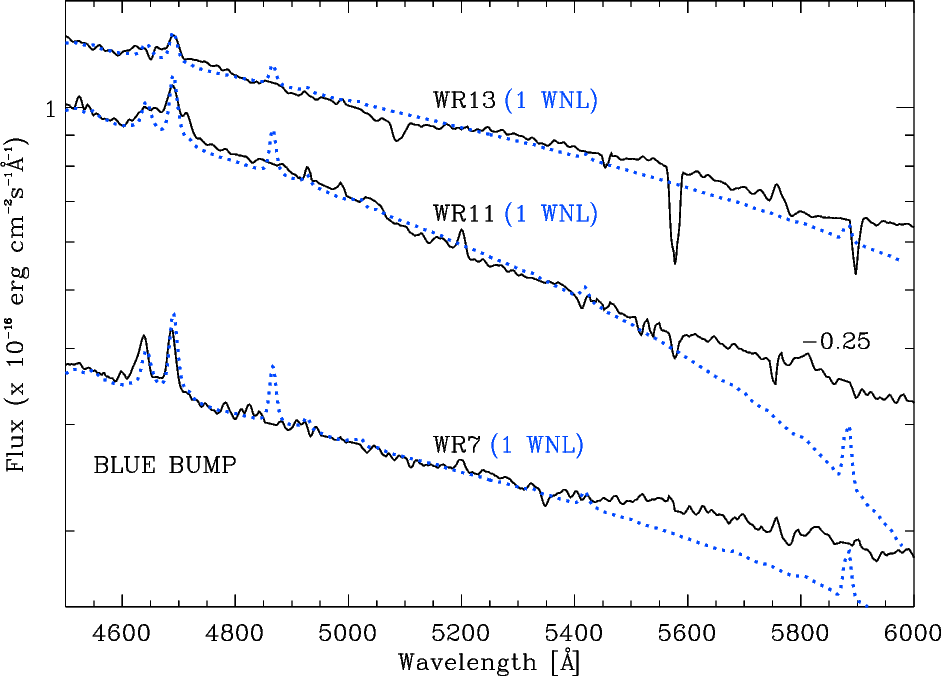}
\includegraphics[width=8.5truecm]{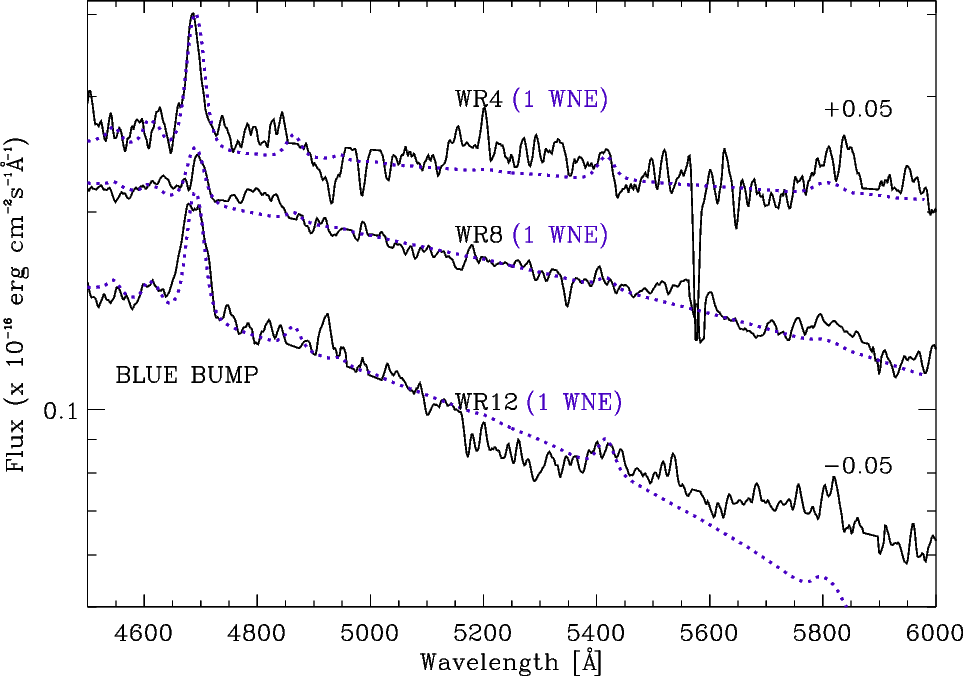} \\
\includegraphics[width=8.5truecm]{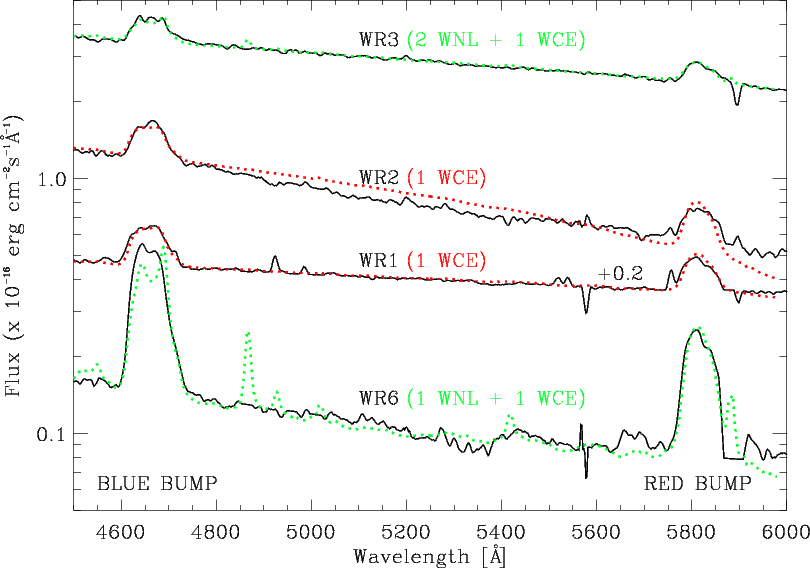}
\includegraphics[width=8.5truecm]{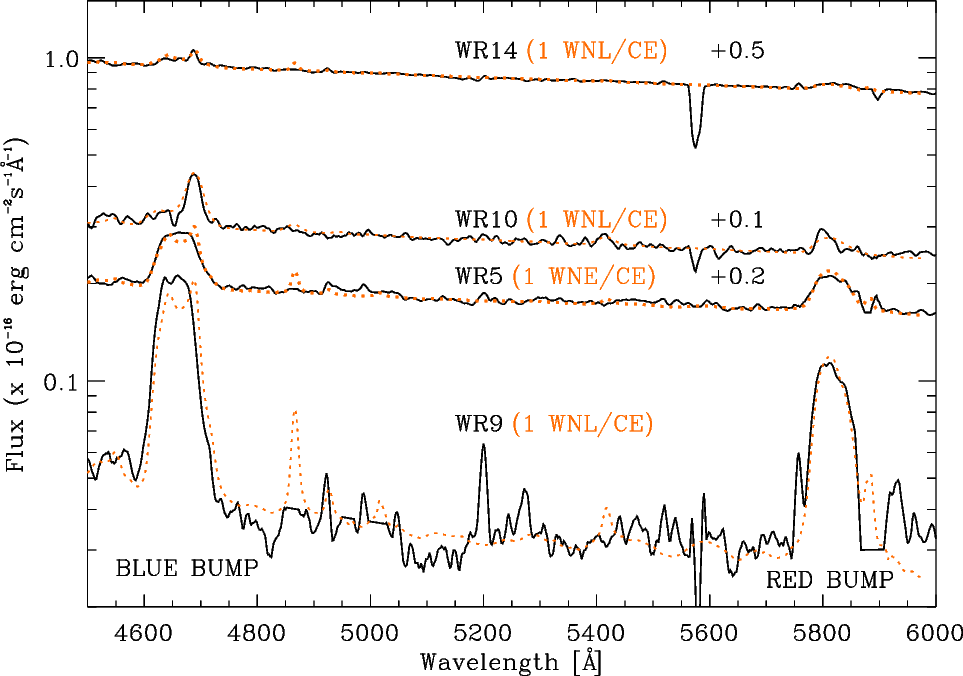}
\par\end{centering}
\caption{\label{fig:temps1}
Observed nebular-free spectrum (solid line) is shown along with the best-fit
template W-R spectrum (colour dotted line). The 
estimated number of W-R stars is indicated in parenthesis. To avoid overlap, 
spectra are shifted by the indicated amount ($-0.25$~dex, +0.05~dex etc.) in a few cases.
The top two panels contain WNL and WNE stars. 
Spectra of our locations containing at least one WCE star are shown in the
bottom-left panel, and the transitional WN/C stars are shown the bottom-right panel.
}
\end{figure*}

The best-fit combination of template spectra is shown superposed on the nebular-free
observed spectra in Figure~\ref{fig:temps1}. The spectra are organized in 4 panels, with the
first two corresponding to WNLs and WNEs, respectively, the next one containing WCEs
in single and multiple stars. The last panel shows the cases that require both
WN and WC with \fwn$+$\fwc$\lesssim$1.

The scaling factor between the template and the observed bump strengths is an indicator of
the number of W-R stars that are required to produce the observed strength of the
bumps in our spectra. The values of \fwn\ and \fwc\ lie in the range of 0.3 to 1.7,
and 0.5 to 1.5, respectively. These values indicate
that our spectra, though passing the cluster centres, require at the most a couple
of W-R stars. The derived non-integer scaling factor is not unexpected, given the large 
dispersion in the strength of W-R features in the spectra used to obtain the template 
spectra.

\citet{2006Crowther} find that the 
strength of the red bump varies by 48\% in different WCE stars, whereas the
blue bump strength varies by as much as 60\% and 71\% in WNLs and WNEs, respectively.
Thus, the scaling factor for a single WCE star could lie anywhere between 0.5--1.5,
for a single WNL star between 0.4--1.6, and for a single WNE star between 0.3--1.7.
We used this criterion to determine the number of WNL, WNE and WCE stars
contributing to a spectrum. 

Before we go ahead determining the number of W-R stars contributing to each
spectrum, we discuss the effect of metallicity differences between
the template and our spectra. Metallicity of ionized nebulae in M81 has 
been studied in detail by \citet{2016Arellano}, who found a mean value of
$12+\log({\rm \frac{O}{H}})=8.58\pm0.06$ with a gradient as small as 0.01~dex~kpc$^{-1}$.
The templates used for fitting correspond to that of the LMC, which has a 
metallicity around a third of the values for our regions. 
According to \citet{2006Crowther},
the W-R spectrum at higher metallicities is expected to be similar to that
of the LMC template used in our work, with the strength of the bumps increasing
with increasing metallicities. Thus, the bumps in M81 spectra are expected to
be marginally stronger than those in the template spectra. As a consequence,
single star \fwc\ and \fwn\ values could be marginally higher than 1 for 
the W-R stars in M81.

Among the 8 cases where we required only one template, 3 are WNLs (\fwn=
0.5, 1.0 and 1.2), 3 WNEs (\fwn=0.4, 0.7 and 1.4) and 2 WCEs (\fwc=0.5 and 1.0). 
Using the criteria discussed above for quantification of number of
W-R stars, the derived scaling factors suggest 1 star in all these cases, even
after taking into consideration differences in metallicity between the
template and M81.
Among the remaining 6 spectra, WR3 and WR6 are multiple systems
requiring at least 3 and 2 W-Rs, respectively, of at least one WNL and WCE sub-types.
The remaining 4 stars have \fwc\ values less than
the minimum expected for a single WCE star. In fact the sum of \fwc+\fwn\ 
(0.4, 0.45, 0.6 and 1.18) suggests a single star. All these spectra have strong
red bump, indicating the presence of a WC component, and the blue bump clearly 
contains the \niii\ line, the characteristic signature of a WN component.
Do these 4 cases represent multiple systems containing sub-luminous W-R stars?
As discussed above, the metallicity difference between the template and M81, if at
all, is expected to increase the strengths of the bumps, and hence systematically
sub-luminous W-R stars are not expected due to metallicity differences.
However, there exist sub-luminous WN stars such as WR24, 
a WN6ha star in the Milky Way, whose bump strength is an order of magnitude 
lower as compared to other stars of the same sub-type \citep{2007Crowther}. 
These are extreme cases, and hence it is statistically improbable 
that 4 of our objects contain multiple stars of extreme WN and WC types.

Are these 4 objects transitional W-R stars?   
\cite{1989Conti} defined a sample of 8 Galactic W-R stars that have prominent
spectral lines corresponding to both WN and WC stars, which they called as
transitional objects (WN/Cs). Stellar evolutionary tracks of 
Geneva \citep{2003Meynet} with rotation
produce these stars as those stars evolving from WN to WC subtype, with this 
transitional phase lasting for $\sim4$\% of the W-R lifetime. Thus, they are 
expected to comprise only $\sim4$\% of our sample.
In fact, the first transitional object in M31 is reported only recently after 
the discovery of more than 150 W-R stars \citep{2016Shara}. In this context, it is extremely 
rare to have 4 transitional types in our sample of 14 W-Rs. In the next section, we 
discuss the photometric and spectroscopic properties of these stars in
order to find out whether they could be two independent stars within our
slit or they are most likely to be transitional objects.

\section{Discussion}

Two of our 14 W-R locations require multiple W-R stars. Among the remaining 12,
eight are clear cases of single stars (3 WNLs, 3 WNEs and 2 WCEs). The remaining
4 cases require features from both WN and WC templates, but not strong enough 
to suggest 2 independent WN and WC stars.
We now analyse all our stars in diagrams involving the line strengths, magnitudes
and FWHM. In particular, we investigate whether the 4 
stars that require both WN and WC templates could be of transitional type,
classified as WN/C type, could have contribution from two or
more stars, at least one each being WN and WC.
 
W-R bumps have asymmetric form, with contributions from multiple lines.
Multiple Gaussian fitting has been used by \citet{2008Brinchmann} with 
considerable success to obtain line strengths of individual components. 
We here intend to obtain the total strength of the bumps, and
hence used simple numerical integration (the command stroke {\it e} 
of {\it IRAF/splot}) covering the blue and red bumps. When the red bump 
is not detected we determined an upper limit by calculating the 3~$\sigma$ 
fluxes over a width of 80~\AA. We also measured the width of the blue and
red bumps. The red bump is seen as a single broad line, to which we fitted 
a Gaussian profile to determine the FWHM. On the other hand, sub-structures are seen in
all of our blue bumps. We fitted the Gaussian profile to the brightest of these
sub-structures, which in most cases coincided with the rest wavelength
of \heiiwr, to obtain the FWHM of the blue bump.

\begin{figure}
\begin{centering}
\includegraphics[width=8.0truecm]{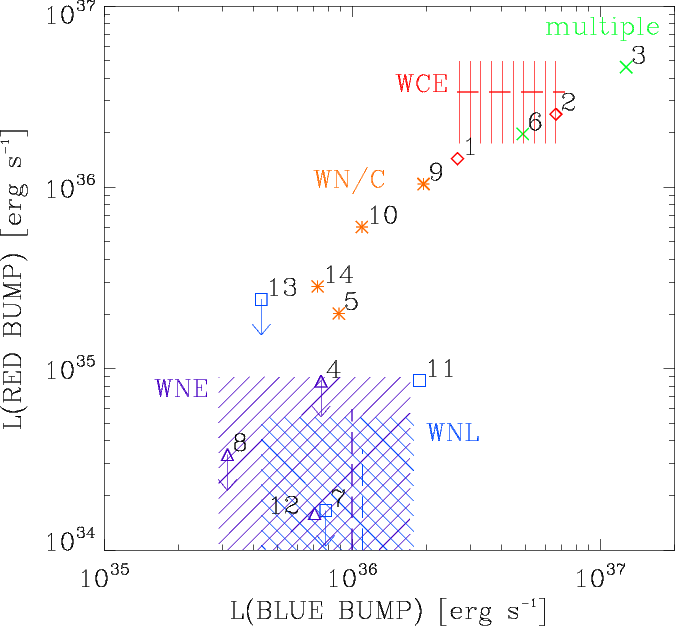}
\par\end{centering}
\caption{\label{fig:new1}
The luminosity of the red bump against that of the blue bump for our sample
of W-R objects. Stars of same subtype are shown by a common symbol and colour. 
3$\sigma$ upper limits are shown when the red bump is not detected. The hatched
areas show the values expected for the WNL, WNE and WCE templates of single stars.
}
\end{figure}

In Figure~\ref{fig:new1}, we plot the luminosity of the red bump against that of the
blue bump. Different W-R subtypes are shown by different symbols. 
The typical values of the bump strength as well as the dispersion over
these values for WNL, WNE and WCE stars are also indicated. 
The plotted ranges for the blue bump in WN and WC stars, and the red bump in WC stars 
are based on the values in Table~1 and Table~2 of \citet{2006Crowther}. 
Considering that a range of the red bump strength (or an upper limit) in WN
stars is not given in these tables, we used the WN template spectra to measure 
their strength. 
In order to plot an upper limit, we assumed a dispersion of 50\% over the measured values,
which are in the same range as the values tabulated by \citet{2006Crowther}.

As expected, the classified WNL, WNE and WCE stars have both their blue and
red bump luminosities within the range of values for the corresponding sub-types. 
W-R locations with multiple W-Rs (WR3 and WR6) occupy a zone in the range
for WCE stars or above it. These systems contain at least one WCE star,
which have both their bumps much brighter than those in WN stars, 
and hence dominate the combined luminosities. On the other hand,
for the transitional types, the observed blue bump luminosities are in
the range expected for WN stars, but they have clearly excess red
bump luminosities compared to that of WN stars. At the same time,
the observed red bump luminosities are lower than the values expected
for the WC type. Moreover, the transitional stars follow a diagonal
line where the red bump luminosity increases proportionately with
the  blue bump luminosity. It is hard to imagine a combination of 
multiple W-R stars of WN and WC types contriving to produce the
observed trend. Any such combination would require WC stars that are underluminous
by as much as by a factor of 10 in their red bump strength.

\begin{figure}
\begin{centering}
\includegraphics[width=8.0truecm]{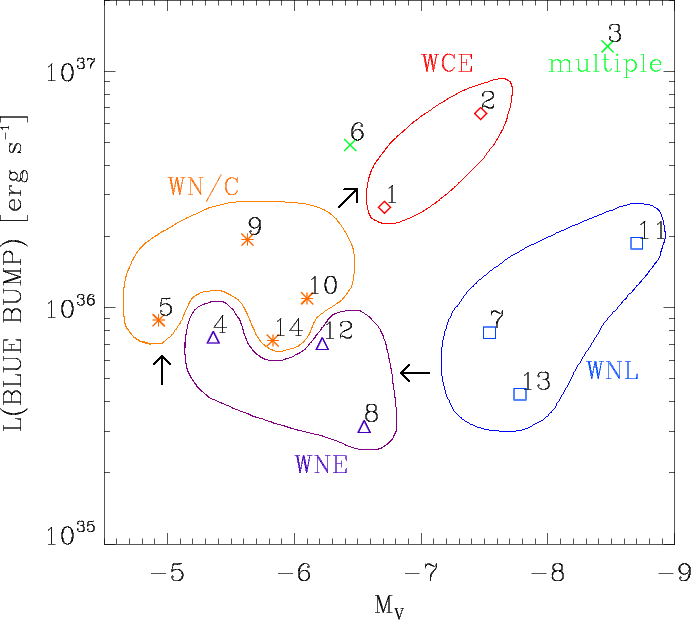}
\par\end{centering}
\caption{\label{fig:new2}
The blue bump luminosity of our W-R stars against the magnitude 
of the star in the HST image with which we associate the observed features.
Stars of the same subtype are enclosed in contours. Arrows indicate
the currently accepted evolutionary scenario in which massive O stars
first appear as W-R stars of subtype WNL, going through WNE, and transitional
WN/C phase before reaching the WC type.
}
\end{figure}

We now plot in Figure~\ref{fig:new2}
the blue bump luminosity of our W-R stars against the magnitude 
of the star in the HST image with which we associate the observed features.
Same symbols as in the previous figure have been retained. All stars of
a given sub-type group together, which are shown by roughly drawing 
contours enclosing the stars of that sub-type. The WNL sub-type, which
is the earliest stage of a W-R phase, occupies the right-bottom part.
The WNE stars, which are the descendants of WNL stars, occupy the 
left-bottom part. On the other hand, the WCE stars, which denote the
last phase of a W-R star, occupy the upper part of the diagram. 
We have indicated by arrows this evolutionary scenario, first proposed
by \citet{1976Conti}. Modern stellar evolutionary models \citep{2003Meynet}
reproduce the observed properties of W-R stars through the Conti scenario.
These models, in addition, were able to reproduce the WN/C stars
as those undergoing transition from WNE to WC phase. In our figure,
these WN/C stars occupy a region between WNE and WC stars, further
illustrating that these stars are transitional stars rather than 
multiple systems, which occupy the top-most part of our diagram.
It may be noted that our WN/C stars are among the faintest in our
sample again hinting that it is unlikely that it is a multiple system of stars.

\begin{figure}
\begin{centering}
\includegraphics[width=8.0truecm]{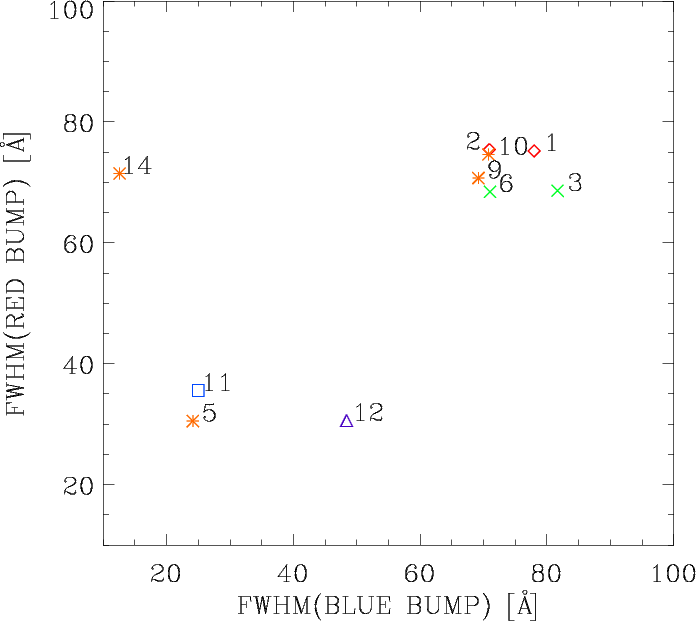}
\includegraphics[width=8.0truecm]{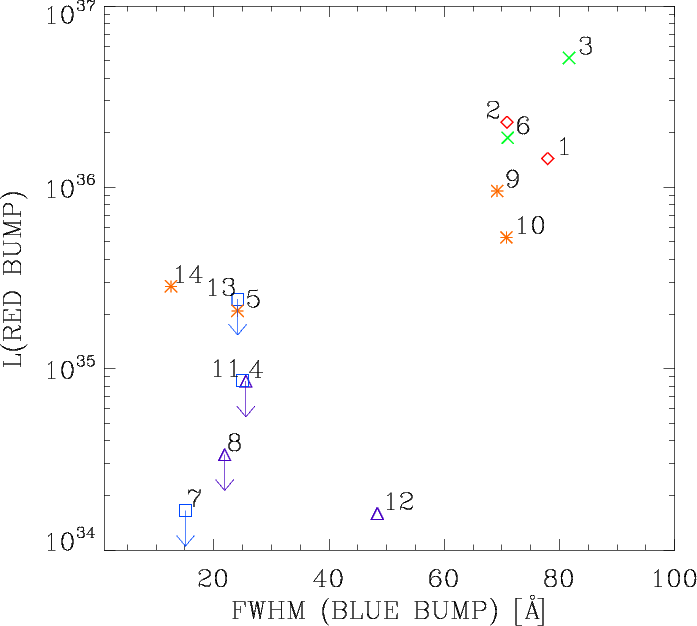}
\par\end{centering}
\caption{\label{fig:new4}
The FWHM of the red bump against the FWHM of the blue bump (top)
and red bump luminosity against the FWHM of the blue bump (bottom).
}
\end{figure}

One of the characteristic signatures of a WC star is its broad bumps with the
red bump having FWHMs $>$50~\AA\ \citep[see e. g. Figure 4 in][]{2006Crowther}.
In Figure~\ref{fig:new4}, we show the FWHM of the red bump against the FWHM of the blue bump.
As expected, WCE stars
occupy the top-right corner of the plot having values between 70$<$FWHM$<$100~\AA. 
Two of our transitional stars, and multiple systems, also occupy this corner. 
One of the transitional stars (WR5) has FWHM$<30$~\AA\ for both the bumps. This
latter value is the typical value of FWHM in WNs \citep[see e. g. Figure 2 in][]{2006Crowther}. 
We illustrate this in the bottom panel
by plotting the observed strength of the red bump (or an upper limit when the red bump is not detected) against
the FWHM of the blue bump. It is interesting to note that there is
only one object (WR12) with measured FWHM between 30--70~\AA. 
In the transitional object WR5, both blue and red bumps have their FWHM of a WN star. 
On the other hand, in WR14, another transitional object, the red bump is as broad
as that of WCs, and the blue bump has FWHM typical of WNs. The other two transitional
stars have both their bumps broad like in WCs.
Hence, mixed properties of WR5, WR9, WR10 and WR14 suggest that they are more
likely to be transitional objects of type WN/C rather than multiple system
containing a WN and a WC star.

\begin{table}
\begin{center}
\caption{\label{tab:table3} Classification of the W-R stars of our sample.}
\begin{tabular}{lrrrr}
\hline
ID & $f_{\rm WNL}$ & $f_{\rm WNE}$ & $f_{\rm WCE}$ & $Class$ \\
(1) & (2) & (3) & (4) & (5) \\
\hline
WR1 & --  & --  & 0.5 &1WCE     \\
WR2 & --  & --  & 1.0 &1WCE     \\
WR3 & 1.9 & --  & 1.5 &2WNL+1WCE\\
WR4 & --  & 1.4 & --  &1WNE     \\
WR5 & 1.1 & --  & 0.1 &1WNE/CE  \\
WR6 & 0.9 & --  & 0.6 &1WNL+1WCE\\
WR7 & 0.5 & --  & --  &1WNL     \\
WR8 & --  & 0.4 & --  &1WNE     \\
WR9 & 0.3 & --  & 0.3 &1WNL/CE  \\
WR10& 0.2 & --  & 0.2 &1WNL/CE  \\
WR11& 1.0 & --  & --  &1WNL     \\
WR12& --  & 0.7 & --  &1WNE     \\
WR13& 1.2 & --  & --  &1WNL     \\
WR14& 0.4 & --  & 0.1 &1WNL/CE  \\

\hline
\end{tabular}\\
(1) W-R identification number;
(2) Factor by which WNL template has been multiplied to reproduce the observed features;
(3) Factor by which WNE template has been multiplied to reproduce the observed features;
(4) Factor by which WCE template has been multiplied to reproduce the observed features;
(5) W-R classification of each star;
\end{center}
\end{table}

\section{Conclusions}

In this work, we reported the discovery of 14 locations in M81 where
we found spectral signatures for the presence of W-R stars.
All the locations were serendipitously discovered along the GTC long-slit 
and MOS spectra of targeted compact stellar clusters. At each location,
we identified a candidate object in the HST image that is responsible 
for the W-R features. We analysed our W-R spectra using template spectra of
WN and WC stars to obtain the number and sub-type of W-R star(s) that are
responsible for the strength of the observed features. We find clear case
for multiple W-R stars in two locations, with the remaining 12 locations
requiring only one W-R star. We classify 3 of them as WNLs, 3 as WNEs, and
2 as WCEs. Four stars are found to have their red bumps too strong to be of WN
type, but not strong enough to be of WC star. All the observed properties
of these 4 stars are consistent with them being transitional WN/C stars.
However, 4 out of our sample of 14 is statistically high as compared to the
4\% expected in stellar evolutionary models. A narrow-band imaging survey
to detect W-R bumps of the whole galaxy would be required to address the 
reason for the observed abnormally high fraction of transitional stars in 
our serendipitous sample.

\section*{ACKNOWLEDGEMENTS}

It is a pleasure to thank Antonio Cabrera and the rest of the GTC staff
for their help in carrying out the observations presented in this work and 
also for the support during data reductions.
We thank the Hubble Heritage Team at the Space Telescope Science
Institute for making the M81 images publicly available. 
This work is partly supported by CONACyT (Mexico) research grants 
CB-2010-01-155142-G3 (PI:YDM) and CB-2011-01-167281-F3 (PI:DRG).
VMAGG thanks CONACyT for the research scholar fellowship that is granted to him.
The paper has gained enormously from the suggestions of a referee on an
earlier version.

\end{document}